\documentclass[fleqn,twocolumn,3p,times,procedia]{elsarticle}

\usepackage{ecrc}
\usepackage{graphicx}
\usepackage{subfig}
\usepackage{xcolor} 
\usepackage{tikz}
\usepackage{pifont}
\usepackage{stfloats}
\usepackage{balance}
\usepackage{lipsum}
\usepackage{comment}
\usepackage{tikz}
\usepackage{multirow}
\usepackage{pifont}
\usepackage{array,threeparttable}
\usepackage{url}

\volume{00}

\firstpage{1}

\runauth{C.V. Radhakrishnan et al.}

\jid{procs}

\CopyrightLine{2022}{Published by Elsevier Ltd.}

\usepackage{amssymb}
\usepackage{amsmath}
\usepackage{multicol}
\usepackage{algorithm}
\usepackage{algpseudocode}
\usepackage[figuresright]{rotating}
\usepackage{bm}

\usepackage{caption}
\usepackage{titlesec}
\titleformat*{\section}{\small \bf}
\titleformat*{\subsection}{\small \em}
\titleformat*{\subsubsection}{\small \em}
\newcommand*\circled[1]{\tikz[baseline=(char.base)]{
    \node[shape=circle,draw,inner sep=1pt, scale=0.8] (char) {#1};}}

\usepackage{fancyhdr}
\fancyheadoffset[RO,EL]{0pt}
\usepackage{geometry}
\begin{document}\small
\biboptions{numbers,sort&compress}
\begin{frontmatter}

\dochead{}

\title{
\begin{flushleft}
{\LARGE Research and Experimental Validation for 3GPP ISAC Channel Modeling Standardization}
\end{flushleft}
}

\author[]{ \leftline {Yuxiang Zhang $^a$, Jianhua Zhang $^*$$^a$, Jiwei Zhang $^a$, Yuanpeng Pei $^a$, Yameng Liu $^a$, Lei Tian $^a$}
\leftline {Tao Jiang $^b$, Guangyi Liu $^b$}}

\address{ \leftline {$^a$State Key Lab of Networking and Switching Technology, Beijing University of Posts and Telecommunications, Beijing, China}
\leftline {$^b$China Mobile Research Institute Beijing, Beijing 100053, China}
}

\begin{abstract}
Integrated Sensing and Communication (ISAC) is considered a key technology in 6G networks. An accurate sensing channel model is crucial for the design and sensing performance evaluation of ISAC systems. The widely used Geometry-Based Stochastic Model (GBSM), typically applied in standardized channel modeling, mainly focuses on the statistical fading characteristics of the channel. However, it fails to capture the characteristics of targets in ISAC systems, such as their positions and velocities, as well as the impact of the targets on the background. To address this issue, this paper proposes an extended GBSM (E-GBSM) sensing channel model that incorporates newly discovered channel characteristics into a unified modeling framework. In this framework, the sensing channel is divided into target and background channels. For the target channel, the model introduces a concatenated modeling approach, while for the background channel, a parameter called the power control factor is introduced to assess impact of the target on the background channel, making the modeling framework applicable to both mono-static and bi-static sensing modes. To validate the proposed model's effectiveness, measurements of target and background channels are conducted in both indoor and outdoor scenarios, covering various sensing targets such as metal plates, reconfigurable intelligent surfaces, human bodies, UAVs, and vehicles. The experimental results provide important theoretical support and empirical data for the standardization of ISAC channel modeling.
\end{abstract}


\end{frontmatter}

\section{Introduction}
Integrated Sensing and Communication (ISAC) is regarded as a key cutting-edge technology to enable ubiquitous sensing capabilities within 6G communication systems\cite{ref:ITU愿景}. By integrating communication and sensing into a single device, ISAC leverages a shared spectrum and hardware platform, significantly enhancing the efficiency of spectrum and energy resource utilization. ISAC can support various emerging application scenarios, including Unmanned Aerial Vehicle (UAV) surveillance, human identification, vehicle platooning, etc.\cite{ref:ISAC综述}.

In wireless communication, the channel serves as the propagation medium for electromagnetic waves between the transmitter (Tx) and receiver (Rx), whose physical properties are essential for the design and application of communication systems\cite{ref:需要的信道模型综述2}.
In traditional communication systems, the focus of research has been on characterizing the statistical fading properties of channels, which determine the ergodic channel capacity\cite{ref:统计簇调研}. The Geometry-Based Stochastic Model (GBSM) is the most popular channel models at present, where the Channel Impulse Responses (CIRs) are determined by the geometric positions of scatterers and the probability density function (PDF) of channel statistical characteristic parameters\cite{jiang2024high}. These parameters are derived from extensive field measurements, enabling the GBSM model to effectively capture the statistical fading properties of channels across various real-world scenarios.
In ISAC systems, in addition to characterizing the statistical fading properties, it is necessary to precisely model the position, velocity, and echo properties of the sensing target (ST) as well as the sensing modes of the channel (mono-static and bi-static). Consequently, the GBSM described in 3rd Generation Partnership Project (3GPP) Technical Report 38.901\cite{ref:TR38.901}, which serves as the 5G stochastic standard channel model, is not directly applicable to ISAC channel modeling due to its inability to accommodate the properties of ISAC channels.

To address this, in December 2023, 3GPP established an industry specification group focused on ISAC sensing channel modeling\cite{ref:TR22.137}. Several meetings have been held\cite{ref:116次会议, ref:116b次会议, ref:117次会议, ref:118次会议, ref:118b次会议}, with general agreement to use 3GPP TR 38.901 as a starting point to initiate the standardization. At the \#116th meeting, an overall framework for sensing channel modeling was approved, in which the sensing channel consists of a target and a background channel. The target channel includes all propagation paths influenced by the target, affected by the target's position, velocity, and echo properties, while the background channel primarily includes background clutter caused by environmental scatterers other than the target\cite{ref:雷达相关调研}. In the standardization process, extensive research and discussions have been conducted separately for the target and the background channel. However, many challenges still remain.

\subsection{Related works}
For the target channel, it is necessary to model the Tx-ST-Rx propagation link, whereas traditional communication channels only require modeling of the Tx-Rx link without considering the ST\cite{ref:裴元鹏news}. \cite{ref:北交综述和簇替换}\cite{ref:Oppo提案} modify the target channel using an non-concatenated method based on the statistical clusters generated by GBSM. They assume the possible locations of the ST and find the corresponding links in the GBSM, then make appropriate modifications to generate new links. Finally, they replace the old links in the GBSM to obtain the modified sensing channel model. The computational complexity of this non-concatenated method is relatively high, and the mapped position of the ST may not correspond to a physically existing location, which presents a challenge for accurately evaluating the sensing performance. \cite{ref:理论分析级联建模}\cite{ref:背景调研3} model the target channel based on the concatenated method. They consider scenarios where the propagation links include line-of-sight (LOS) paths and modeled the Tx-ST and ST-Rx links separately, then evaluated the sensing performance through the concatenated Tx-ST-Rx link. These assumptions all assume ideal channel propagation, but there are complex scatterers in scenarios such as indoor hotspots or dense urban areas, which lead to a higher power ratio of Non-Line-of-Sight (NLOS) paths\cite{ref:裴元鹏news}\cite{ref:李3}. In these cases, it is more appropriate to model the NLOS conditions. Therefore, how to model the Tx-ST-Rx link within complex scattering scenarios still requires substantial field measurements and researches.

For the background channel, it is primarily related to the sensing mode, while its correlation with the target channel also needs to be considered.
On the one hand, the sensing mode is determined by the relative position of the Tx and Rx. Like traditional communication systems, locations of the Tx and Rx are not co-located in bi-static sensing mode, thus the modeling method in 3GPP TR 38.901\cite{ref:TR38.901} may be used to generate background channel. However, in the mono-static sensing mode, the Tx and Rx are at the same location. Although channel models for mono-static sensing mode exist in traditional radar systems\cite{ref:单站与雷达关系的调研}, they are primarily applied in the aviation field for monitoring tasks and are not suitable for most ISAC scenarios.
On the other hand, the correlation between background and target channels also impacts the algorithm design of ISAC systems. \cite{ref:共享簇调研3} designs an ISAC system based on this correlation and theoretically demonstrates that it can enhance the sensing performance of the ISAC system. Currently, most ISAC channel models treat the background and target channels independently\cite{ref:共享簇调研5}\cite{ref:共享簇调研6}. However, \cite{ref:共享簇调研2} and \cite{ref:共享簇调研4} found through experiments that some shared scatterers exist between these two channels. It may be possible to model these shared scatterers to describe the correlation between the background and target channels. Therefore, how to improve the channel model for the mono-static mode and whether the correlation between the background and target channels needs to be characterized still requires extensive field measurements and research.

\subsection{Contributions}
To address these issues, we propose corresponding channel models for the target and background channels, complemented by extensive field measurements. Specifically, the contributions of this paper are as follows:

\begin{itemize}
    \item We propose an Extended Geometry-Based Stochastic Model (E-GBSM) for ISAC channel, which encompasses both target and background channel modeling. For the target channel, we introduce a concatenated channel modeling method. Building upon this, we explore the correlation between the target and background channels and develop a background channel model that is applicable to both mono-static and bi-static sensing modes.

    \item To validate the concatenation property of the target channel, we conducted field measurements in various indoor and outdoor scenarios involving metal plates, RIS, UAVs, vehicles, and humans. Our findings indicate that the CIR of the Tx-ST-Rx links can be constructed by convolving the CIRs of the Tx-ST and ST-Rx links. Additionally, we observed that NLOS paths account for a significant portion of the power in these links. Therefore, it is essential to incorporate NLOS path modeling in our proposed model.
    
    \item To investigate the correlation between the background and target channels, we conducted measurements in various indoor and outdoor scenarios involving halls and humans in both mono-static and bi-static sensing modes. We introduced a new Power Control Factor (PCF) to characterize the impact of the ST on the background channel. Our findings reveal that the PCF can be generated statistically, which can then be utilized for the joint design of background and target channel models.
\end{itemize}

The remainder of this paper is organized as follows:
Section 2 introduces the existing 3D GBSM used in 5G standards and highlights the unique properties of ISAC channels compared to 5G communication channels through comparative analysis. Based on this, an E-GBSM is proposed to describe these properties.
Section 3 provides a detailed description of the equipment used for channel measurements and the scenario setup.
Section 4 presents the data processing methods and conducts an in-depth analysis of various ISAC channel properties.
Section 5 concludes the paper and discusses potential future research directions.

\section{ISAC Channel Modeling Framework}
This section first introduces the 3D GBSM model used in current 5G standards. The new properties that need to be considered in the sensing channel of ISAC systems are then introduced. These include accurately modeling NLOS paths in the target channel, addressing the concatenation properties of the target channel, establishing unified modeling of the background channel under various sensing modes, and understanding the coupling effects between background and target channels. Finally, building on the 5G standard channel model, we propose an E-GBSM to capture these unique properties of the sensing channel.

\subsection{The existing 3D GBSM channel modeling method in 5G standards}
According to the theoretical illustration of the 5G 3D GBSM model in \cite{ref:3DMIMO的JSAC}\cite{ref:需要的信道模型综述1}, as shown in Fig.~\ref{fig:ISAC整体特性大图}(a), assuming that Tx and Rx are equipped with arrays of \( S \) and \( U \) antennas, respectively, the channel coefficient from the \( s \)-th antenna of the transmitter to the \( u \)-th antenna of the receiver is expressed as
\begin{equation}
    h_{u,s}^{\text{5G}}(t,\tau) = \sum_{n=1}^{N} \sum_{m=1}^{M} h_{u,s,n,m}^{\text{5G}}(t) \cdot \delta(\tau - \tau_{n,m}),
\end{equation}
where \( N \) and \( M \) represent the number of clusters and the number of paths within each cluster, respectively. The channel coefficient \( h_{u,s,n,m}^{\text{5G}}(t,\tau) \) for the \( m \)-th ray within the \( n \)-th cluster from the \( s \)-th antenna of the transmitter to the \( u \)-th antenna of the receiver is given by
\begin{align}
    h_{u,s,n,m}^{\text{5G}}(t) &= \sqrt{\frac{P_{n}}{M}} 
    \mathbf{F}^{\text{Rx} \, \text{T}}_{u,n,m}
    \mathbf{CPM}_{n,m}
    \mathbf{F}^{\text{Tx}}_{s,n,m} \cdot \exp \left( \text{j} 2 \pi f_{n,m} t \right) \notag \\ 
    & \quad \cdot \exp \left( \text{j} \frac{2 \pi}{\lambda} \left(\mathbf{r}_{\text{Rx},n,m}^{\text{T}} \cdot \mathbf{d}_{\text{Rx},u}+\mathbf{r}_{\text{Tx},n,m}^{\text{T}} \cdot \mathbf{d}_{\text{Tx},s}\right) \right),
\label{eq:5G-GBSM模型}
\end{align}
where,
\begin{itemize}
    \item \( (\cdot)^{\text{T}} \) represents the transpose of the matrix, and \( \lambda \) denotes the carrier wavelength.
    
    \item \( P_n \) represents the power of the \( n \)-th cluster, and \( \tau_{n,m} \) is the delay of the \( m \)-th ray in the \( n \)-th cluster.
    
    \item \( \mathbf{r}_{\text{Rx},n,m}, \mathbf{r}_{\text{Tx},n,m} \) denote the unit direction vectors corresponding to the angle of arrival and angle of departure, respectively. Correspondingly, \( \mathbf{d}_{\text{Rx},u}, \mathbf{d}_{\text{Tx},s} \) represent the position vectors of the receiving antenna \( u \) and transmitting antenna \( s \).

    \item $f_{n,m}$ represents the Doppler frequency.

    \item \( \mathbf{F}^{\text{Rx}}_{u,n,m} \) and \( \mathbf{F}^{\text{Tx}}_{s,n,m} \) represent the radiation patterns of the receiving and transmitting antennas at the corresponding polarization and angle, respectively, with the expressions as follows:
\begin{equation}
    \mathbf{F}^{\text{Rx}}_{u,n,m} = \left[
        \begin{array}{c}
        F_{\text{Rx},u}^{\theta} \left( \mathbf{\Gamma}_{\text{Rx},n,m} \right) \\
        F_{\text{Rx},u}^{\phi} \left( \mathbf{\Gamma}_{\text{Rx},n,m} \right)
        \end{array}
    \right],
    \mathbf{F}^{\text{Tx}}_{s,n,m} = \left[
        \begin{array}{c}
        F_{\text{Tx},s}^{\theta} \left( \mathbf{\Gamma}_{\text{Tx},n,m} \right) \\
        F_{\text{Tx},s}^{\phi} \left( \mathbf{\Gamma}_{\text{Tx},n,m} \right)
        \end{array}
    \right],
\end{equation}
where
\begin{itemize}
    \item \( F_{\text{Rx},u}^{\theta}, F_{\text{Rx},u}^{\phi}, F_{\text{Tx},s}^{\theta}, F_{\text{Tx},s}^{\phi} \) represent the vertical polarization component and horizontal polarization component for the radiation pattern of the receiving antenna \( u \) and the transmitting antenna \( s \), respectively.

    \item \( \mathbf{\Gamma}_{\text{Rx},n,m} \) is the angle of arrival (AoA) of \( (n,m)^{th} \) path. It includes the azimuth angle of arrival (AAoA) \(\phi_{\text{Rx},n,m}\) and the elevation angle of arrival (EAoA) \(\theta_{\text{Rx},n,m}\) in 3D space. Similarly, the angle of departure (AoD) \( \mathbf{\Gamma}_{\text{Tx},n,m} \) represents the \( (n,m)^{th} \) path, where the azimuth angle of departure (AAoD) \(\phi_{\text{Tx},n,m}\) and the elevation angle of departure (EAoD) \(\theta_{\text{Tx},n,m}\) are defined.
\end{itemize}

\item \( \mathbf{CPM}_{n,m} \) is the cross-polarization matrix, with expressions as follows:
\begin{equation}
	\mathbf{CPM}_{n,m} = \left[
	\begin{array}{cc}
		\exp \left( \textnormal{j} \Phi_{n,m}^{\theta \theta} \right) & \sqrt{\kappa_{n,m}^{-1}} \exp \left( \textnormal{j} \Phi_{n,m}^{\theta \phi} \right) \\
		\sqrt{\kappa_{n,m}^{-1}} \exp \left( \textnormal{j} \Phi_{n,m}^{\phi \theta} \right) & \exp \left( \textnormal{j} \Phi_{n,m}^{\phi \phi} \right)
	\end{array}
	\right],
\end{equation}

where \( \Phi_{n,m}^{p_1 p_2}, p_1, p_2 \in \{\theta, \phi\} \) denotes the initial phase of the \( (n,m) \)-th ray. The angles on the right-hand side represent the polarization from the transmitting antenna \( s \) and the receiving antenna \( u \). \( \kappa_{n,m} \) represents the cross-polarization ratio (XPR) of this ray.
\end{itemize}

\subsection{New features of the ISAC Channel}
\begin{figure*}[ht]
    \centering
    \subfloat[]{\includegraphics[width=0.75\columnwidth]{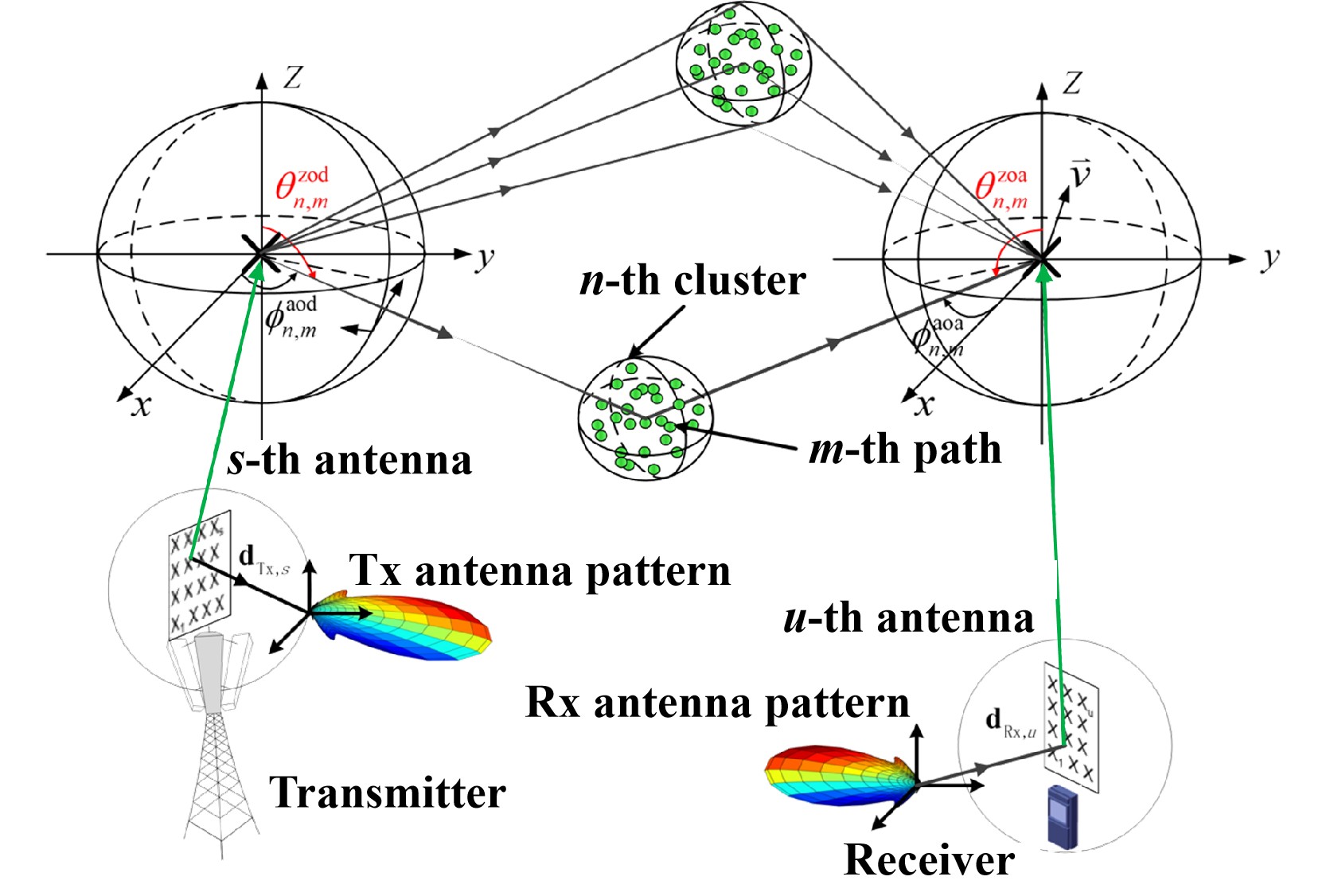}}
    \hfill
    \subfloat[]{\includegraphics[width=1.25\columnwidth]{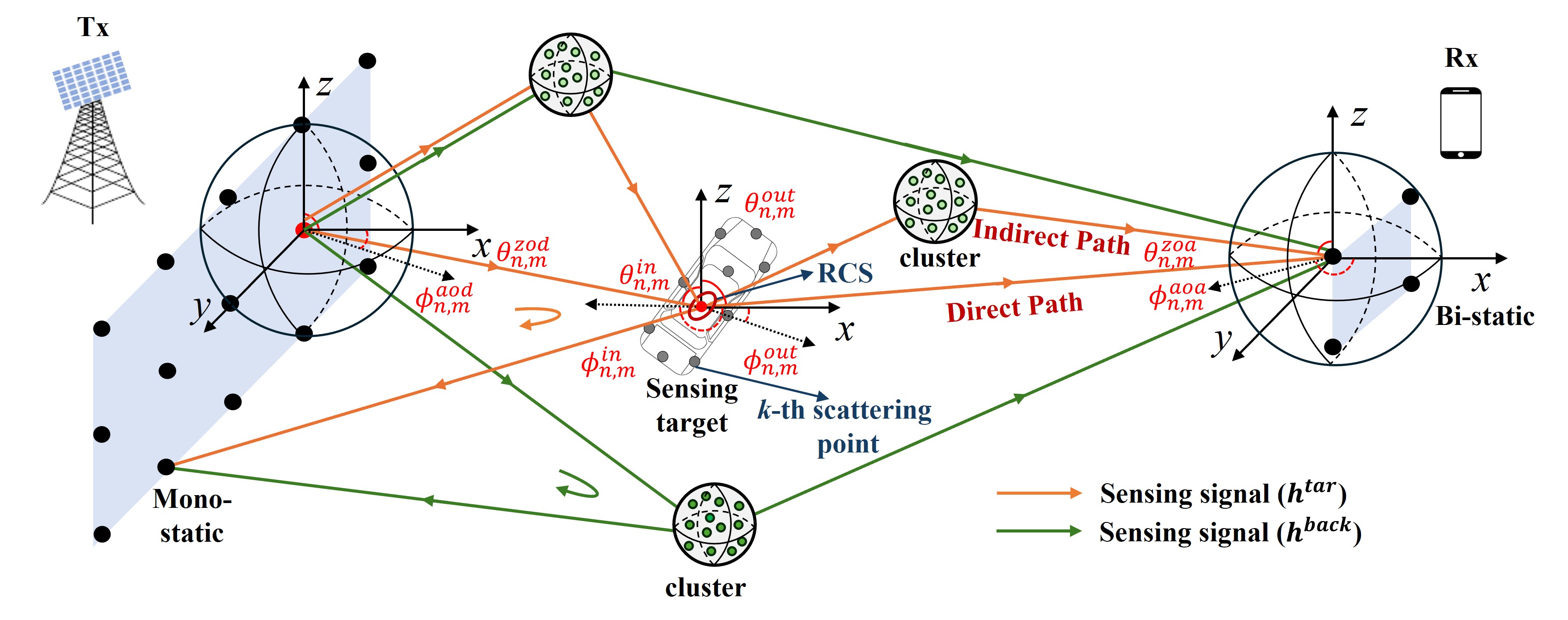}}
    \caption{Schematic diagram of (a) the 5G 3D GBSM channel model\cite{ref:3DMIMO的JSAC}, (b) the 6G ISAC channel model based on extended-GBSM\cite{ref:裴元鹏news}.}
    \label{fig:ISAC整体特性大图}
\end{figure*}

According to the consensus reached at the \#116 meeting, the sensing channel is divided into two parts: the target channel and the background channel. The target channel includes all propagation paths influenced by the target, represented by yellow lines in Fig.~\ref{fig:ISAC整体特性大图} (b). The background channel comprises paths not affected by the target, such as environmental clutter, shown in green lines in Fig.~\ref{fig:ISAC整体特性大图} (b).

\textit{Target Channel.} The target channel requires consideration of modeling and concatenation for the Tx-ST and ST-Rx links. As shown in Fig.~\ref{fig:ISAC整体特性大图} (b), the propagation paths in the target channel can be categorized into two types: direct paths (DPs) and indirect paths (IDPs). DPs refer to signals that travel along a LOS path from Tx to the ST and then directly to Rx via another LOS path. IDPs, on the other hand, involve interactions not only with the target object but also with other environmental objects, potentially experiencing one or more reflections, scatterings, or diffraction, represented as scattering clusters in Fig.~\ref{fig:ISAC整体特性大图} (b). Comparing Fig.~\ref{fig:ISAC整体特性大图} (a) and (b), we can observe that IDPs undergo environmental scattering similar to that in communication channels, allowing the Tx-ST and ST-Rx links to be modeled using a 3D GBSM approach. Additionally, when concatenating the links, the target’s RCS at different angles must also be considered.

\textit{Background Channel.} Background channel modeling should account for differences between bi-static and mono-static sensing modes. In bi-static sensing, particularly between base stations and user terminals, the Tx and Rx positions resemble those of conventional communication channels. Therefore, existing 5G communication channel modeling methods can be referenced, specifically for properties such as path loss, path delay, and angles, in line with current standards. On the other hand, for mono-static sensing mode, where the Tx and Rx are co-located, the existing 5G standard does not yet define this type of channel, requiring further research and measurement experiments to establish references and support.

\textit{Correlation between Background and Target Channels.} As previously mentioned, due to the shared propagation environment in ISAC systems, a correlation exists between the target and background channels. This correlation may arise from the target’s obstruction of existing paths in the background channel or from scatterers in the environment participating in both target and background channel propagation.

\subsection{An extended GBSM for ISAC channel modeling standardization}
To provide a unified description of the new characteristics of target and background channels in ISAC channel modeling, an E-GBSM model based on the 5G 3D GBSM framework is proposed.

\refstepcounter{equation} 
\newcounter{mainEqCounter}
\setcounter{mainEqCounter}{\value{equation}}
    \begin{subequations}
    \renewcommand{\theequation}{\arabic{mainEqCounter}-\arabic{equation}}

    \begin{align}
        \Tilde{h}^{\text{sen}}_{u,s}(t, \tau) &= \sqrt{PL^{\text{sen}}} \cdot h^{\text{sen}}_{u,s}(t, \tau) \notag \\
        &= \sum_{k=1}^K \sqrt{PL^{\text{tar}}_k} \cdot h_{u,s,k}^{\text{tar}}(t, \tau) + \sqrt{PL^{\text{back}}} \cdot h_{u,s}^{\text{back}}(t, \tau),
        \label{eq:感知信道模型}
    \end{align}

    \begin{align}
        h_{u,s,k}^{\text{tar}}(t, \tau) &= \sum_{n_1,m_1}^{N_1,M_1} \sum_{n_2,m_2}^{N_2,M_2} \sqrt{\frac{P_{n_1,m_1}P_{n_2,m_2}}{n_1n_2}} 
        \mathbf{F}^{\text{Rx} \, \text{T}}_{u,n_2,m_2}
        \mathbf{CPM}_{n_2,m_2} \notag \\
        & \quad \cdot \mathbf{CPM}_{k} \sqrt{\sigma_k(\mathbf{\Gamma}_{\text{out},n_2,m_2},\mathbf{\Gamma}_{\text{in},n_1,m_1)}}
        \mathbf{CPM}_{n_1,m_1}
        \mathbf{F}^{\text{Tx}}_{s,n_1,m_1} \notag \\
        & \quad \cdot \exp \left( \textnormal{j} \frac{2 \pi}{\lambda} \left(\mathbf{r}_{\text{Rx},n_2,m_2}^{\text{T}}\cdot\mathbf{d}_{\text{Rx},u}
        +\mathbf{r}_{\text{out},n_2,m_2}^{\text{T}}\cdot\mathbf{d}_{\text{sp},k} \right) \right) \notag \\
        & \quad \cdot \exp \left( \textnormal{j} \frac{2 \pi}{\lambda} \left(\mathbf{r}_{\text{in},n_1,m_1}^{\text{T}}\cdot\mathbf{d}_{\text{sp},k}
        +\mathbf{r}_{\text{Tx},n_1,m_1}^{\text{T}}\cdot\mathbf{d}_{\text{Rx},u}
        \right) \right) \notag \\
        & \quad \cdot \exp \left( \textnormal{j} 2 \pi f_{n_1,m_1,n_2,m_2} t \right) \delta (\tau - \tau_{n_1,m_1}) * \delta (\tau - \tau_{n_2,m_2}),
        \label{eq:目标信道模型}
    \end{align}

    \begin{align}
        h_{u,s}^{\text{back}}(t, \tau) &= \sum_{n_b,m_b}^{N_b,M_b} \sqrt{\frac{P_{n_b,m_b}}{M_b}} 
        \mathbf{F}^{\text{Rx} \, \text{T}}_{u,n_b,m_b}
        \mathbf{CPM}_{n_b,m_b}
        \mathbf{F}^{\text{Tx}}_{s,n_b,m_b} \notag \\
        & \quad \cdot \exp \left( \text{j} \frac{2 \pi}{\lambda} \left(\mathbf{r}_{\text{Rx},n_b,m_b}^{\text{T}} \cdot \mathbf{d}_{\text{Rx},u}+\mathbf{r}_{\text{Tx},n_b,m_b}^{\text{T}} \cdot \mathbf{d}_{\text{Tx},s}\right) \right) \notag \\
        & \quad \cdot \exp \left( \text{j} 2 \pi f_{n_b,m_b} t \right) \delta (\tau - \tau_{n_b,m_b}),
        \label{eq:背景信道模型}
    \end{align}

    \begin{align}
        PL^{\text{sen}} = \sum_{k=1}^K PL_k^{\text{tar}} + PL^{\text{back}} = \sum_{k=1}^K PL_k^{\text{tar}} + O_{\text{back}} \cdot PL_0^{\text{back}}.
        \label{eq:O因子描述的路损模型}
    \end{align}
    \end{subequations}
    \label{eq:感知信道建模框架}

\addtocounter{equation}{-1}

In a sensing channel, the CIR \(\tilde{h}_{u,s}^{\textnormal{sen}}(t,\tau)\) from the \(s\)-th transmitting antenna to the \(u\)-th receiving antenna can be expressed as the product of path loss \(PL^{\textnormal{sen}}\) and the channel coefficient \(h_{u,s}^{\textnormal{sen}}(t,\tau)\). Additionally, the CIR can be decomposed into the superposition of target and background channel impulse responses, as shown in equation~\eqref{eq:感知信道模型}.

Furthermore, for the target channel, we assume a total of \(K\) scattering points. These scattering points may originate from a large target that requires multi scattering points modeling, or from multiple distinct targets. For the \(k\)-th scattering points, its channel coefficient can be represented by equation~\eqref{eq:目标信道模型}. Most symbols retain the same definitions as in equation~\eqref{eq:5G-GBSM模型}, while the meanings of newly introduced symbols are as follows:
\begin{itemize}
    \item $n_i,m_i,i\in \left\{ 1,2 \right\}$ indicate the identifier of cluster and path in Tx-ST and ST-Rx links.
    
    \item \( \mathbf{\Gamma}_{\text{out},n_2,m_2} \) is the AoD of \( (n_2,m_2)^{th} \) path in ST-Rx link. It includes the AAoD \(\phi_{\text{out},n_2,m_2}\) and the EAoD \(\theta_{\text{out},n_2,m_2}\) in 3D space. Similarly, the AoA \( \mathbf{\Gamma}_{\text{in},n_1,m_1} \) represents the \( (n_1,m_1)^{th} \) path in Tx-ST link, where the AAoA \(\phi_{\text{in},n_1,m_1}\) and the EAoA \(\theta_{\text{in},n_1,m_1}\) are defined.
    
    \item \( \mathbf{r}_{\text{out},n_2,m_2}, \mathbf{r}_{\text{in},n_1,m_1} \) represent the unit direction vectors corresponding to the angles of departure and arrival of the target, respectively. Correspondingly, \( \mathbf{d}_{\text{sp},k} \) represents the position vectors of the $k^{th}$ scattering point of the target.

    \item $\mathbf{CPM}_{k}$ is the cross-polarization matrix, representing the potential polarization twisting effects introduced by the $k^{th}$ scattering point of the target.

    \item $\sigma_k$ represents the RCS of the $k^{th}$ scattering point of the target, which is dependent on the target itself, as well as the angles of arrival and departure.
\end{itemize}
Here, the total power of the Tx-ST-Rx link is equal to the product of the sub-link powers, and the total delay is the sum of the sub-link delays, reflecting its concatenation property. Each scattering point’s scattering power is represented by its RCS, and a CPM is introduced to describe its effect on electromagnetic waves of various polarization combinations.

Considering the similarity between the background channel and traditional communication channels, a similar modeling approach can be used, as shown in equation~\eqref{eq:背景信道模型}.

According to equations~\eqref{eq:目标信道模型} and \eqref{eq:背景信道模型}, the generation processes for the target and background channels are independent. To statistically represent the correlation between the target and background channels, the model introduces a large-scale PCF \(O_{\text{back}}\), which is reflected in large-scale path loss. Based on this, the path loss expression for the sensing channel can be reformulated as shown in equation~\eqref{eq:O因子描述的路损模型}\cite{ref:陈文俊人体}, where \( PL_0^{\textnormal{back}} \) represents the path loss of the background channel in the absence of a target.

\section{ISAC Channel Measurement Description}

\begin{table*}[ht]
\centering
\newcommand{\tabincell}[2]{\begin{tabular}{@{}#1@{}}#2\end{tabular}}
\caption{Measurement configuration}
\renewcommand{\arraystretch}{1.5}
\small
\tabcolsep=0.1cm
\begin{tabular}{c|c|c|c|c|c|c|c}
\hline
\hline
\textbf{Scenarios} & \textbf{\tabincell{c}{Scen1: \\ Metal Plate-InH}} & \textbf{\tabincell{c}{Scen2: \\ RIS-InF}} & \textbf{\tabincell{c}{Scen3: \\ Human-InH}} & \textbf{\tabincell{c}{Scen4: \\ UAV-UMi}} & \textbf{\tabincell{c}{Scen5: \\ Vehicle-Outdoor}} & \textbf{\tabincell{c}{Scen6: \\ Human-Outdoor}} & \textbf{\tabincell{c}{Scen7: \\ Indoor hall}} \\ \hline
Targets type & Metal plate & RIS & Human & UAV & Vehicle & Human & $\varnothing$ \\ \hline
Frequency (GHz) & 28 & 6.9 & 105 & 28 & 26 & 28 & 28 \\ \hline
Symbol rate (Msym/s) & 300 & 200 & 600 & 600 & 300 & 1200 & 500 \\ \hline
Bandwidth (GHz) & 0.6 & 0.4 & 1.2 & 1.2 & 0.6 & 2.4 & 1.0 \\ \hline
Sampling rate (GHz) & 0.6 & 0.4 & 1.2 & 1.2 & 0.6 & 2.4 & 1.0 \\ \hline
Tx / Rx gain (dB) & 25 / 3 & 20 / 20 & 25.5 / 5 & 25 / 25 & 30 / 25 & 25 / 25 & 25 / 25 \\ \hline
Tx / Rx HPBW (deg) & 10.31 / Omni & 15 / 15 & 9.9 / Omni & 10.31 / 10.31 & 5 / 10 & 10.31 / 10.31 & 10.31 / 10.31 \\ \hline
Sensing mode & Bi-static & Bi-static & Bi-static & \tabincell{c}{Mono-static/ \\ Bi-static} & Bi-static & Mono-static & \tabincell{c}{Mono-static/ \\ Bi-static}  \\ \hline
\hline
\end{tabular}
\label{tab:测量汇总表}
\end{table*}

\begin{figure*}[t]
    \centering
    \subfloat[Scen1: Metal Plate-InH]
    {\includegraphics[width=0.45\linewidth]{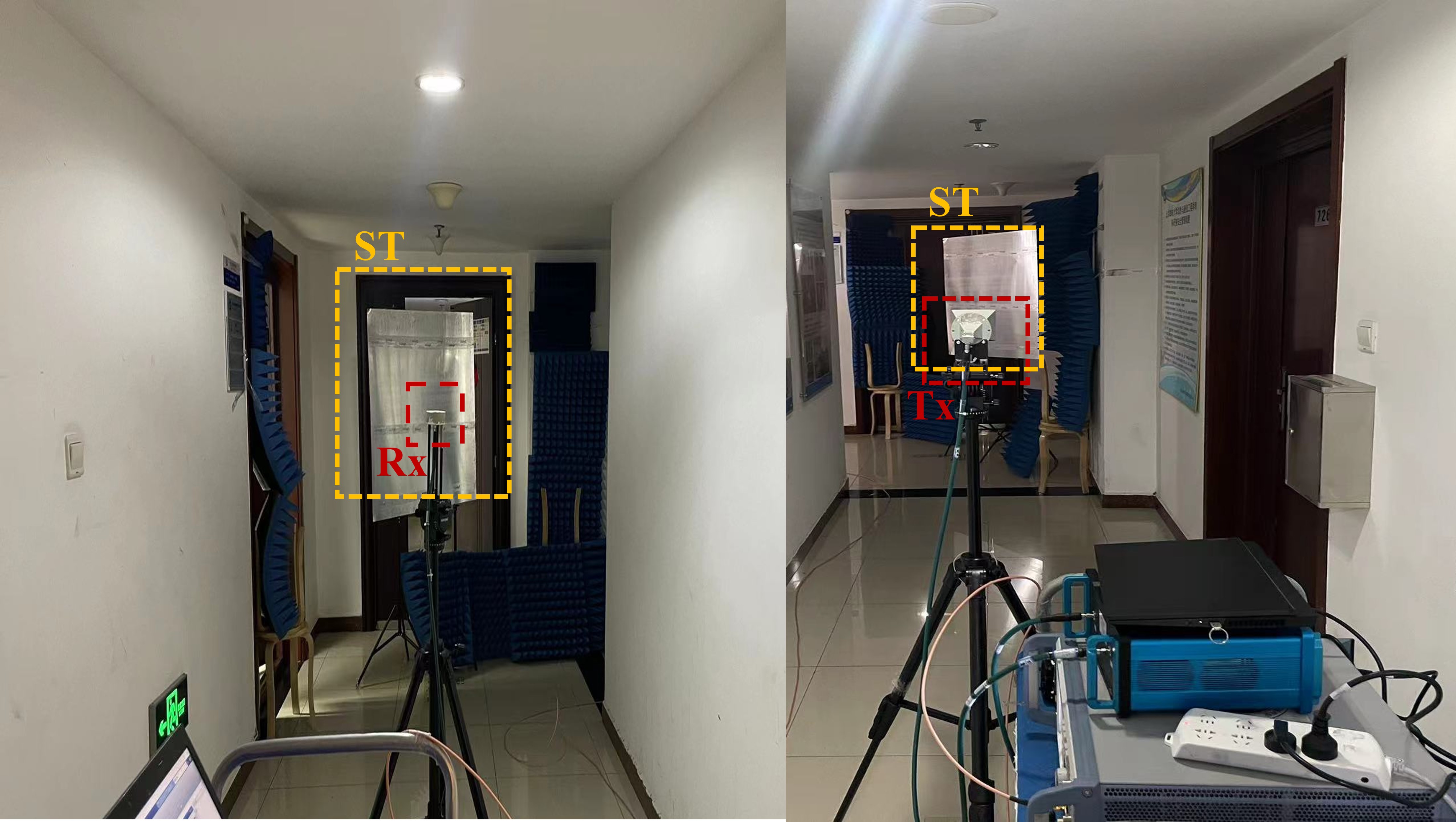}} \hfill
    \subfloat[Scen2: RIS-InF]
    {\includegraphics[width=0.45\linewidth]{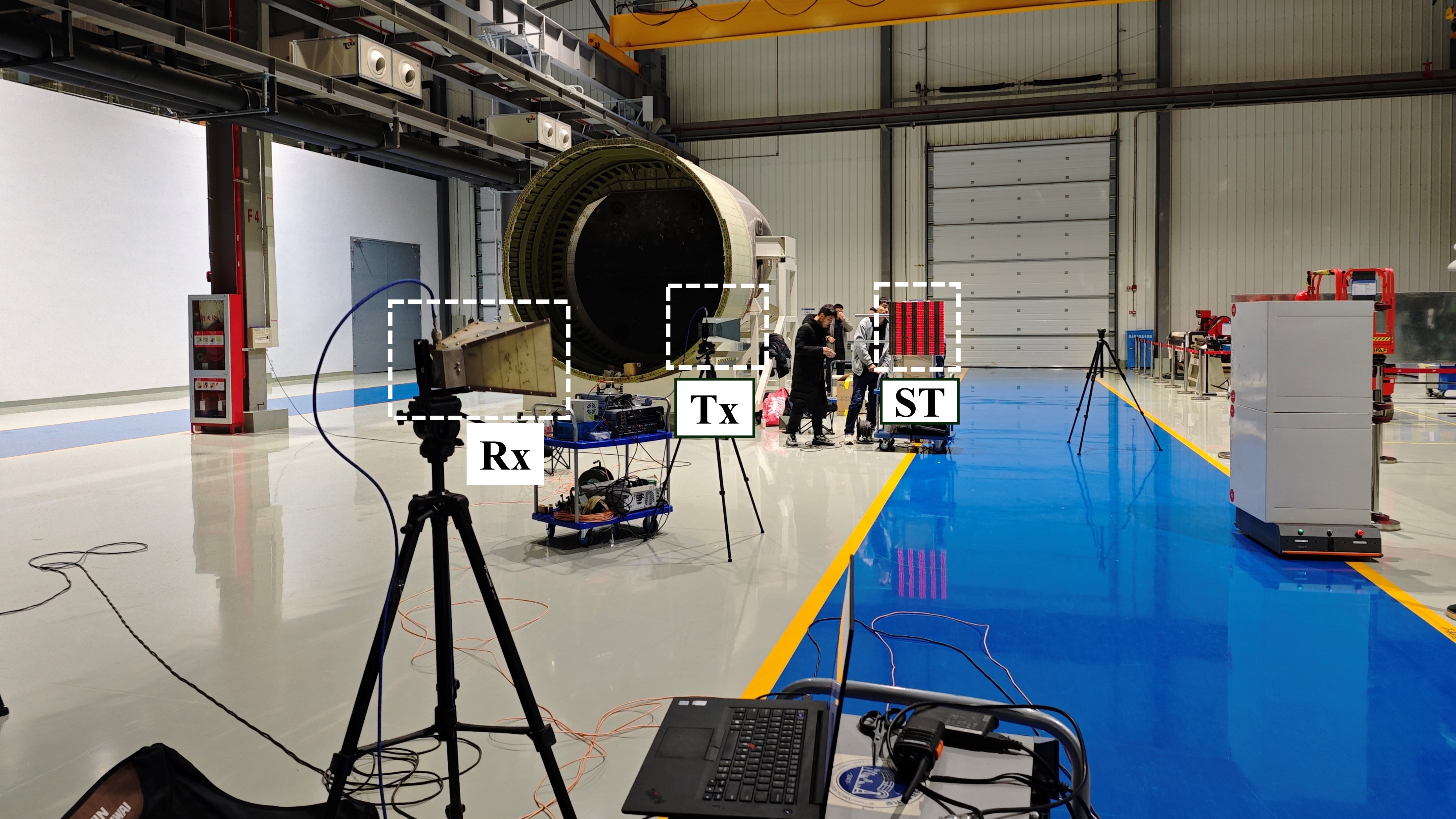}}
    
    \vspace{0.4cm} 
    
    \subfloat[Scen3: Human-InH]
    {\includegraphics[width=0.45\linewidth]{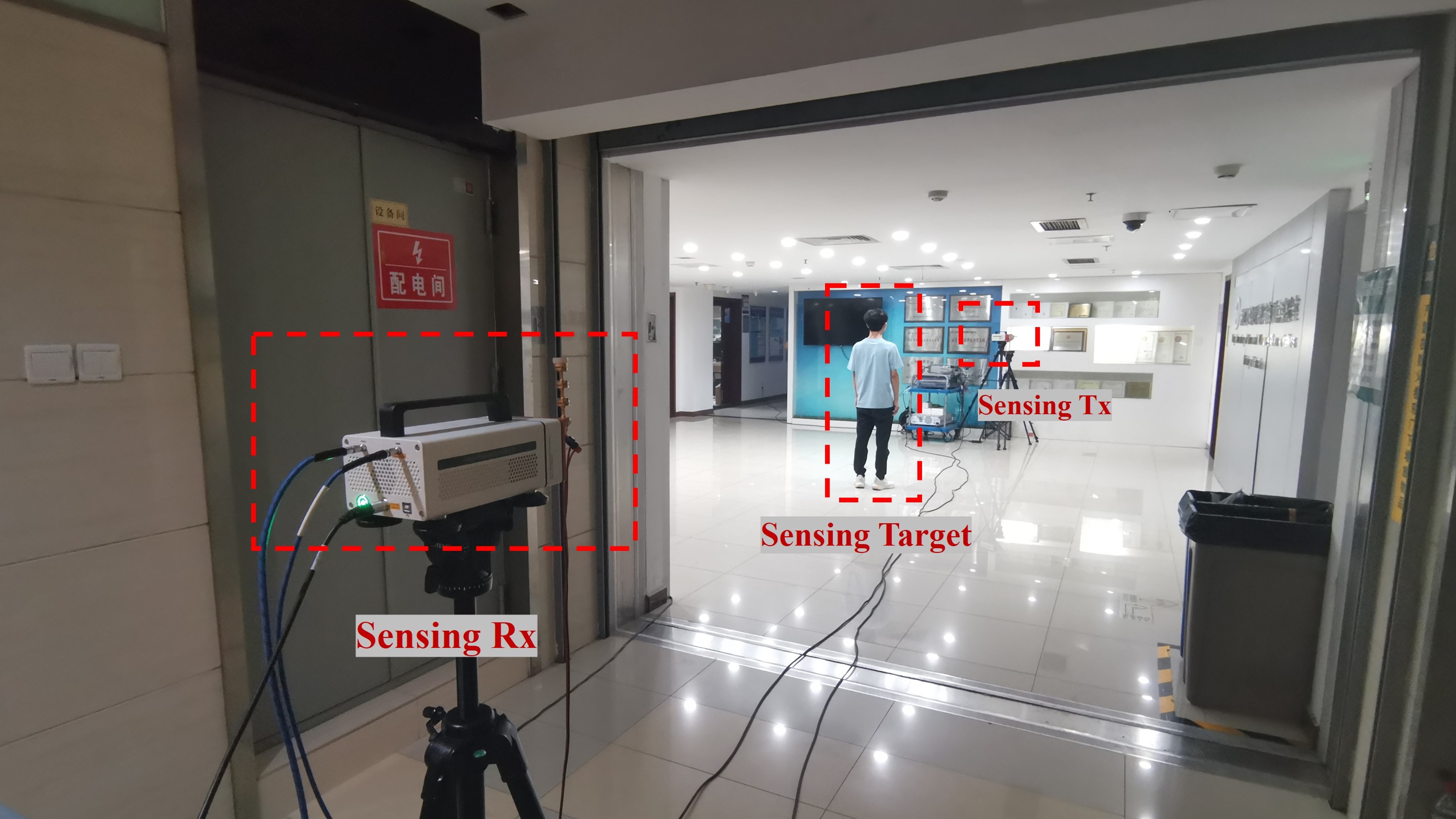}} \hfill
    \subfloat[Scen4: UAV-UMi]
    {\includegraphics[width=0.45\linewidth]{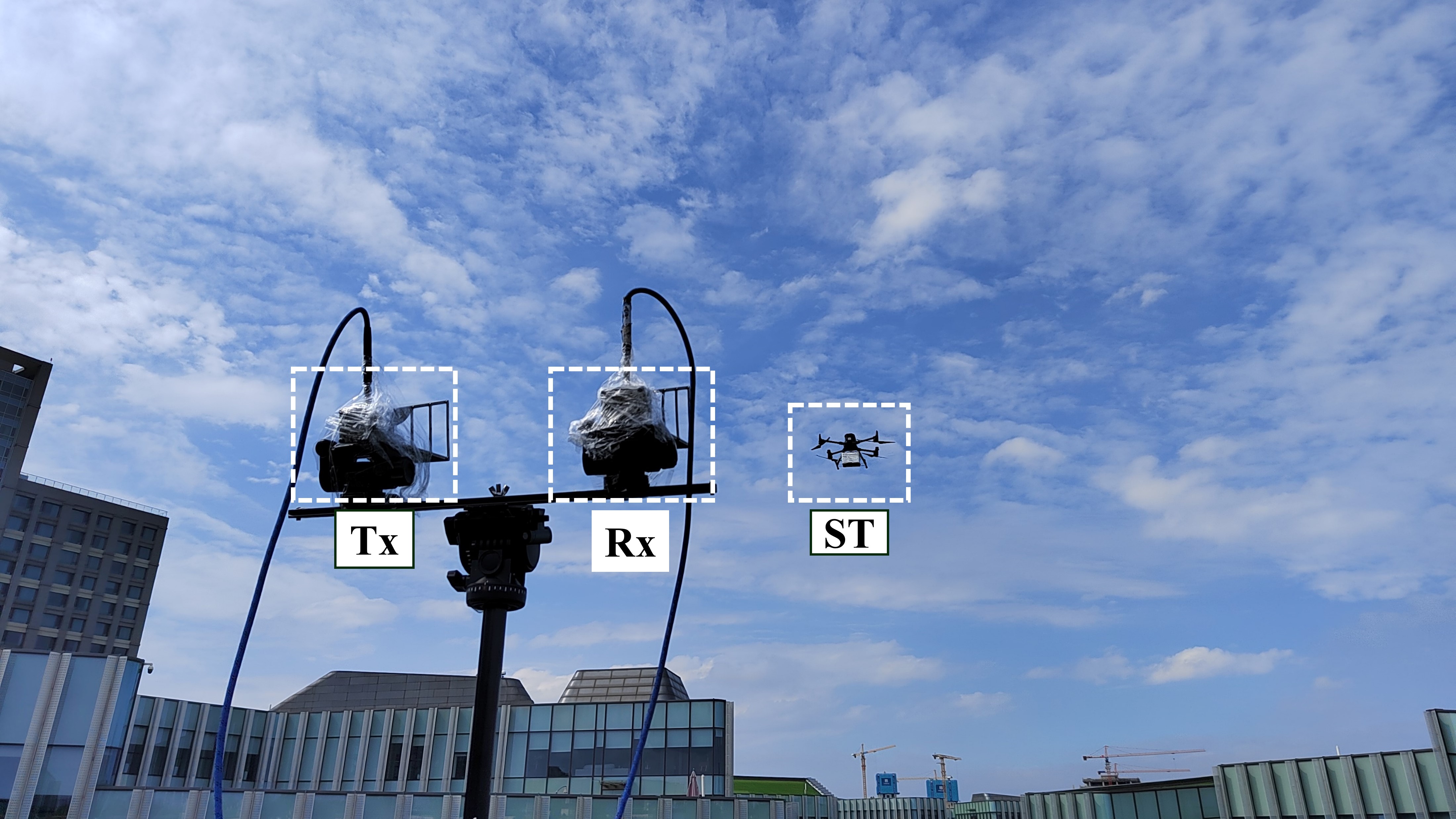}}
    
    \vspace{0.4cm} 
    
    \subfloat[Scen5: Vehicle-Outdoor]{\includegraphics[width=0.3\linewidth]{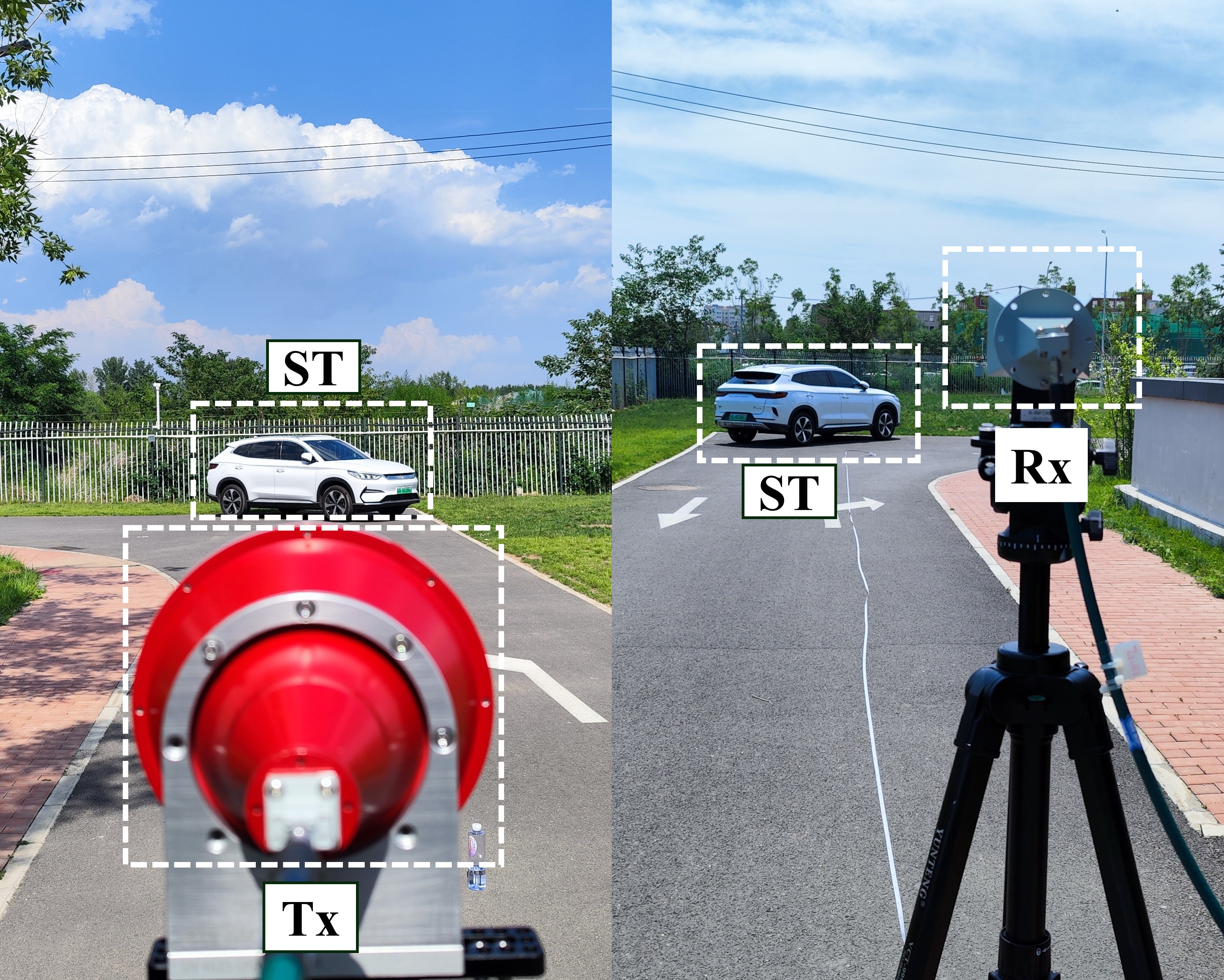}} \hfill
    \subfloat[Scen6: Human-Outdoor]{\includegraphics[width=0.3\linewidth]{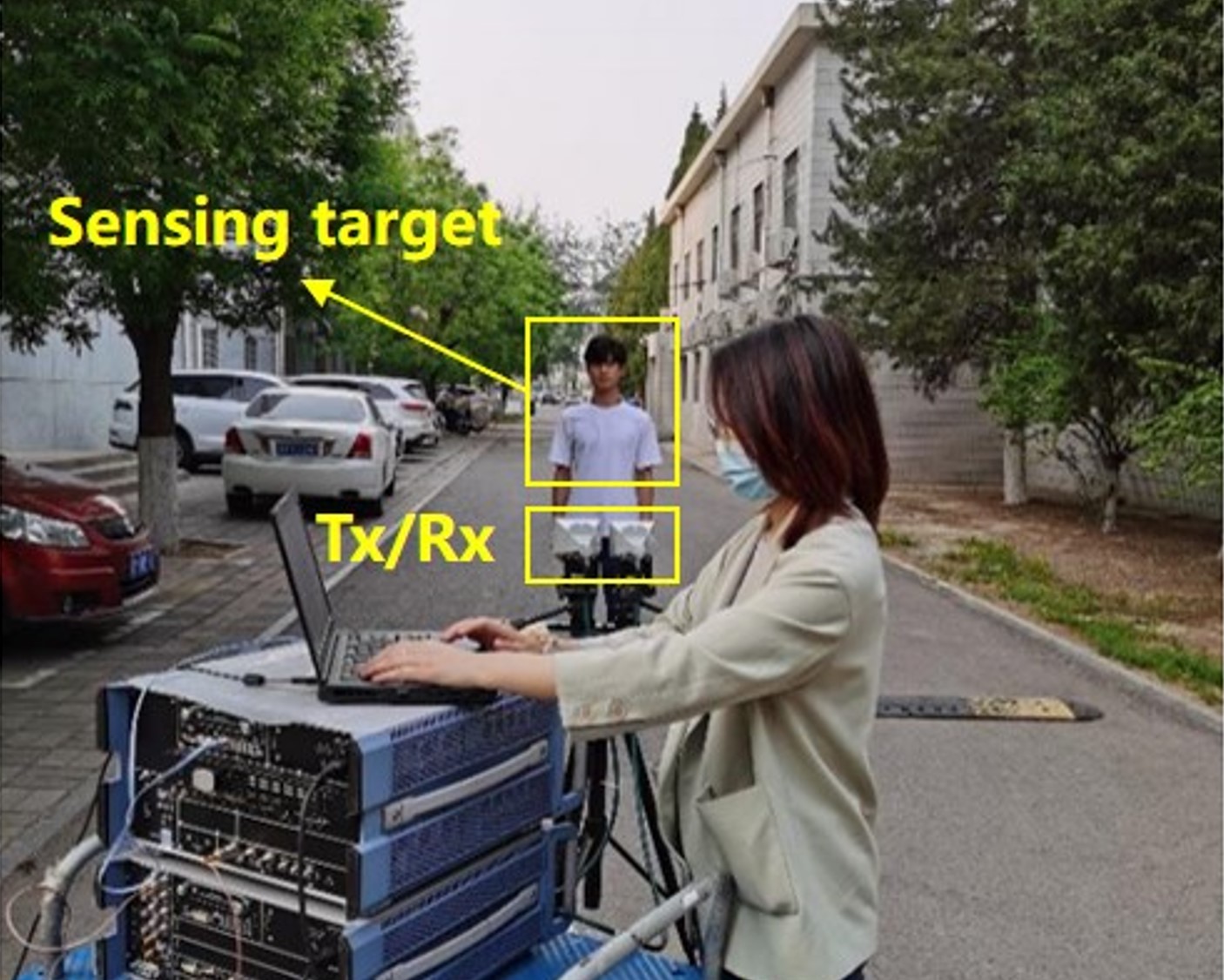}}
    \hfill
    \subfloat[Scen7: Indoor hall]{\includegraphics[width=0.3\linewidth]{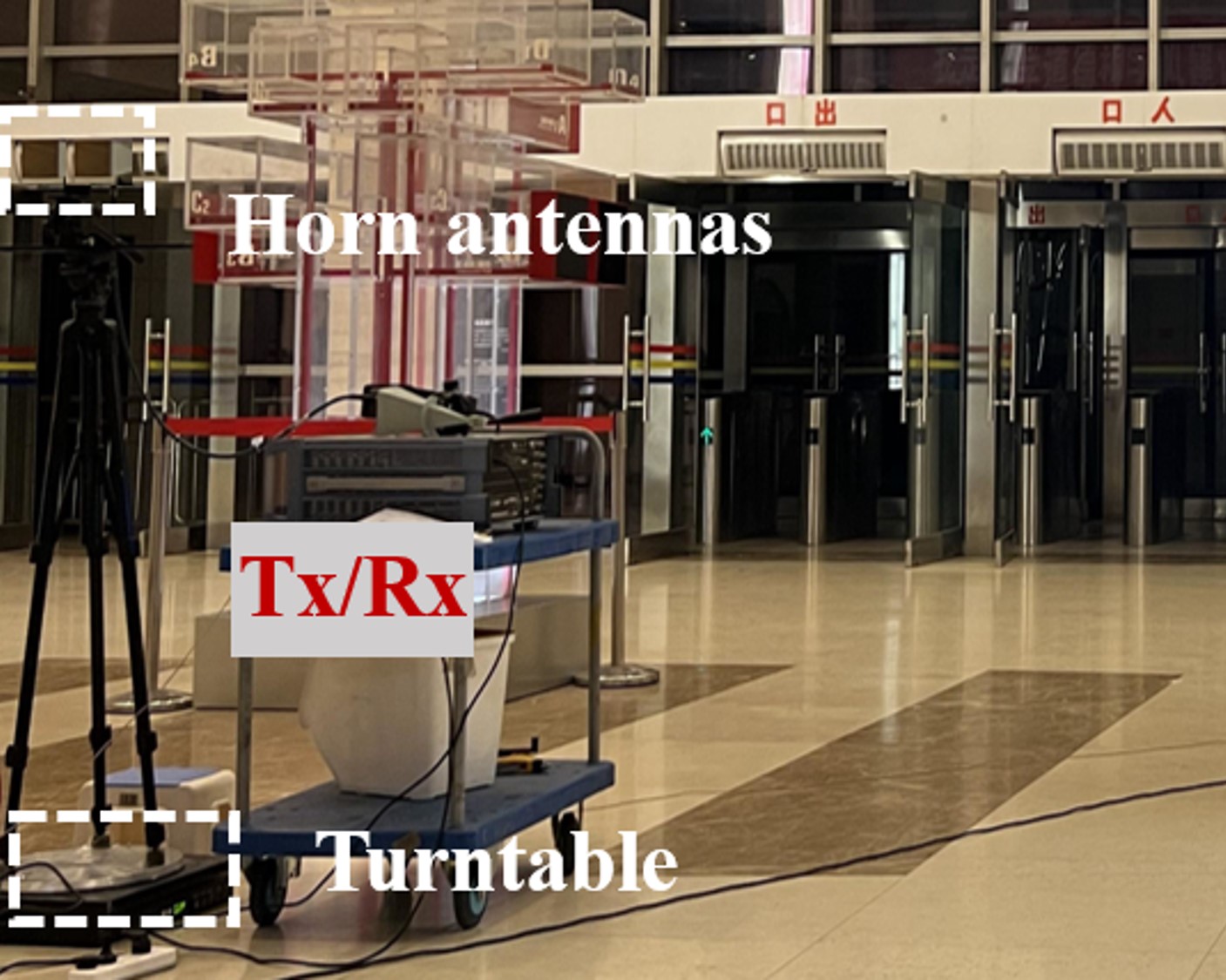}}
    
    \caption{Measurement scenarios.}
    \label{figs:八个场景的测量照片}
\end{figure*}

To verify the reliability of the proposed E-GBSM model in real-world scenarios, a variety of typical sensing targets, including human bodies, drones, and vehicles, are selected, and a series of channel measurements is conducted across multiple environments, such as indoor hotspot (InH), 
indoor factory (InF), and UMi, under different sensing modes, including bi-static and mono-static setups. This section first introduces the time-domain measurement system used and then describes the configuration of each measurement scenario.

\subsection{Measurement system}
In our experiment, we use a wideband channel sounder based on the time-domain sliding correlation principle to extract ISAC channel characteristics. At the transmitter, a vector signal generator produces a pseudo-noise (PN) sequence, which is then modulated using binary phase shift keying to convert it into a signal at the required carrier frequency, and subsequently transmitted into the wireless channel via an antenna. At the receiver, a spectrum analyzer demodulates the signal received by the antenna and captures in-phase and quadrature samples. To enhance the dynamic range of the received signal, a power amplifier is used at the transmitter, and a low-noise amplifier is employed at the receiver. 
Prior to actual measurements, a back-to-back calibration is conducted to eliminate system errors caused by the transmitter, receiver, and connecting cables. By leveraging the excellent autocorrelation properties of the PN sequence, we process the measured data against the calibration data to extract the CIR, thereby enabling precise characterization of the channel properties.

In some scenarios, to study the angular characteristics of the channel, a horn antenna fixed on a turntable is used at either the transmitting or receiving end. The turntable could rotate 360° in 5° increments. By analyzing the collected data, we were able to obtain the multipath propagation characteristics of the channel, including delay, power, and angle, providing data support for the analysis of ISAC system channel characteristics.

\begin{figure}[ht]
    \centering
    \includegraphics[width=\columnwidth]{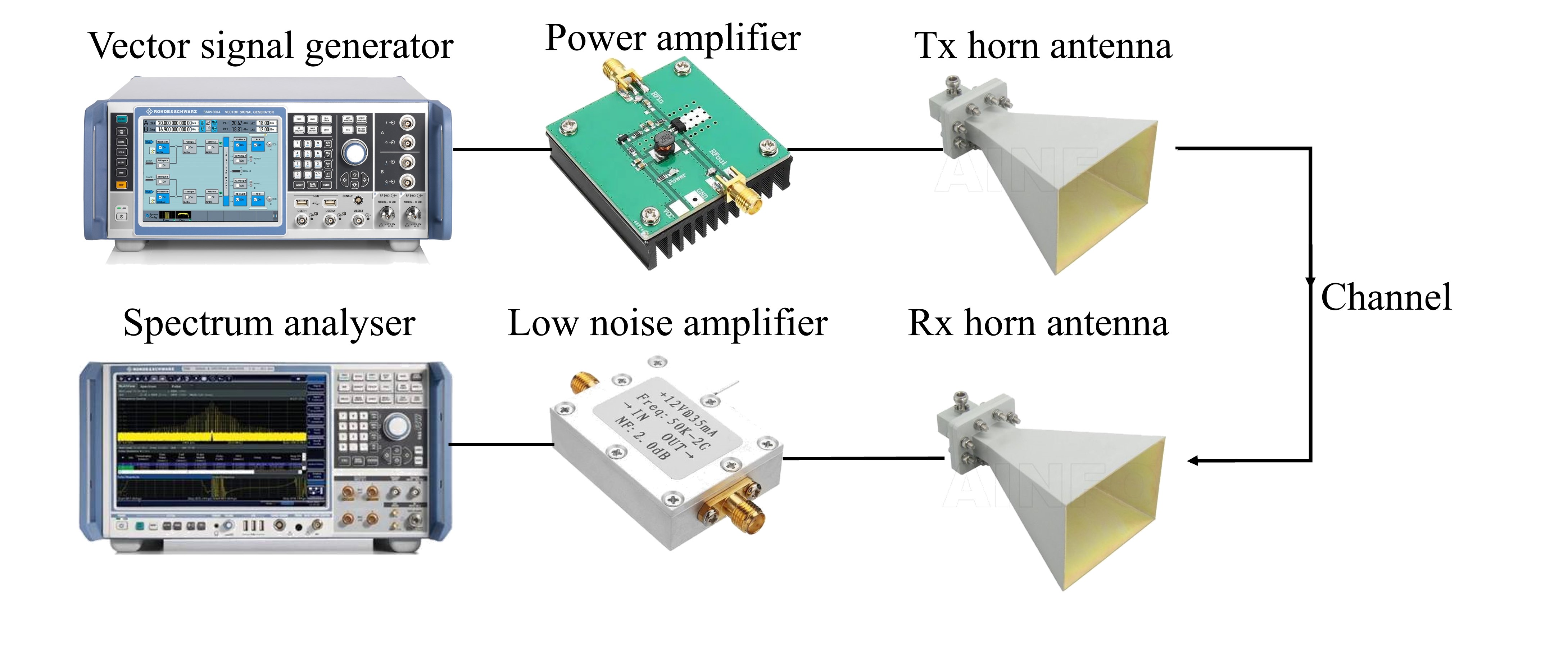}
    \caption{Measurement platform.}
    \label{fig:测量系统框图}
\end{figure}

\subsection{Measurement Scenarios}
\subsubsection{Scen1: Metal Plate-InH}
The measurement takes place in a typical indoor L-shaped corridor. The Tx antenna is an omnidirectional antenna, and the Rx antenna is a horn antenna, both mounted at a height of 1.5 m and placed at opposite ends of the corridor. A smooth metal plate measuring 1 m × 1 m is selected as the target and positioned at a 45° angle at the corridor corner. 

Figure~\ref{figs:八个场景的测量照片}(a) shows the measurement setup. The measurements are divided into Tx-ST-Rx, Tx-ST, and ST-Rx link measurements. In the Tx-ST-Rx link measurements, the initial distances from the Tx and Rx to the target are measured to be 4.45 m and 3 m, respectively. Subsequently, the distances between the Tx/Rx and the target are incrementally increased in predefined steps. Specifically, four measurement points are set up for the Tx along the corridor, while 15 receiving points are spaced at regular intervals for the Rx along the corridor. As a result, a total of 4×15 data sets are collected during the measurements. For the Tx-ST link, the Tx remained stationary, while the Rx is moved to the target’s position, yielding 4 data sets. For the ST-Rx link, the Rx remained stationary, and the Tx is moved to the target’s position, resulting in 15 data sets.

\subsubsection{Scen2: RIS-InF}
The measurement takes place in a typical InF scenario. Both the Tx and Rx antennas are horn antennas, mounted at a height of 1.5 m. The Tx horn antenna always points toward the target, while the Rx horn antenna is mounted on a turntable, allowing it to rotate 360° in 5° increments. A 0.6 m × 0.6 m Reconfigurable Intelligent Surface (RIS) is selected as the target.

Figure~\ref{figs:八个场景的测量照片}(b) shows the measurement setup. The measurements are divided into Tx-ST-Rx, Tx-ST, and ST-Rx link measurements. For the Tx-ST-Rx target link measurement, the horn antenna at the Tx is placed in two positions, marked as Tx1 and Tx2. 
For the Tx-ST link measurement, the Tx antenna is placed in the same positions as in the Tx-ST-Rx link measurements, while the target is replaced with an omnidirectional antenna to measure signal power and delay parameters along the Tx-ST link. For the ST-Rx link, to accurately calculate the RCS of the RIS, it is essential to precisely determine the departure angle of the RIS path. Therefore, we design two sets of experiments to measure the departure angle of the RIS and the angle of arrival at the Rx.
First, the horn antenna is fixed on a rotating platform at the target’s position and rotates in 5° increments over a range of 180° to measure the departure angle from the target. Then, at the Rx side, the horn antenna is similarly fixed on a turntable and rotates continuously through 360°, with measurements taken every 5° to capture the angles of arrival at the Rx, while an omnidirectional antenna is placed at the target position.

\subsubsection{Scen3: Human-InH}
The measurement takes place in a typical InH scenario. Both the Tx and Rx antennas are horn antennas with a height of 1.4 m and a distance of 10 m. The Tx horn antenna is mounted on a turntable, allowing it to rotate from 0° to 180° in 5° increments. A male target with a height of 1.7 m and a shoulder width of 0.4 m is selected for the experiment. To accurately extract path information from the target channel, we obtain background channel data by removing the target from the measurement scenario while keeping all other conditions unchanged. This method is used in all other experiments in this paper to collect background channel data.

Fig.~\ref{figs:八个场景的测量照片}(c) illustrates the measurement setup. During the target channel information collection, we make the human target stand at different measuring points with a static state facing east and orient the Tx horn antenna towards the human target. According to the conditions of Tx-ST and ST-Rx links, the positions of the human target are divided into LOS+LOS and LOS+NLOS conditions.

\subsubsection{Scen4: UAV-UMi}
The experiment is conducted in a typical UMi environment. To recreate the UMi scene, horn antennas for both the Tx and Rx are mounted on rooftops, with a combined height (antenna mount and building) of 17.14 m. An UAV measuring 0.81 m × 0.67 m × 0.43 m, with a cardboard box attached underneath (0.3 m × 0.2 m × 0.15 m) is used to simulate a real-world application.

Fig.~\ref{figs:八个场景的测量照片}(d) illustrates the setup for mono-static measurement, where both the Tx and Rx are at the same location. Considering the impact of UAV altitude on communication performance\cite{ref:李1}\cite{ref:李2}, The UAV hovers at positions of 15°, 0°, and -15° relative to the horizontal plane of the Tx/Rx, with a horizontal distance of 8.95 m from the Tx/Rx. Tx/Rx rotates from 0° to 180° in 5° increments. In the bi-static measurement, where Tx and Rx are placed on opposite sides of the rooftop, separated by 31 m, the UAV hovers between Tx and Rx, maintaining a consistent altitude, with distances of 17.4 m and 17 m from Tx and Rx, respectively. In this case, the Tx antenna is aimed directly at the target, while the Rx rotates from 0° to 180° in 5° increments.

\subsubsection{Scen5: Vehicle-Outdoor}
The measurements are conducted at an L-shaped intersection, an open area surrounded by significant scatterers such as tall buildings, fences, and vegetation, with both the Tx and Rx horn antennas positioned at a height of 1.4 m. An vehicle measuring 4.26 m × 1.78 m × 1.57 m is used as a target.

Fig.~\ref{figs:八个场景的测量照片}(e) illustrates the measurement setup. The experiment consists of two cases: in Case 1, both the Tx and Rx antennas face directly towards the car, while in Case 2, the car's orientation is adjusted to simulate changes during driving, causing the antennas to no longer directly face the car.
In Case 1, where the car is parked at a 45° angle at the corner of the intersection. The transmitter and receiver antennas are placed on opposite sides of the L-shaped intersection, with no LOS path due to building obstructions. During the measurement, the transmitter antenna remains fixed, facing the car, while the receiver antenna rotates from 0° to 180° in 5° increments, with 90° being directly towards the car.
In Case 2, the positions of the antennas and the measurement method remain the same as in Case 1, but the car is repositioned so that there is no longer a LOS path between the receiver antenna and the car. By comparing the results from these two cases, the scattering characteristics of the vehicle in different orientations can be analyzed, providing data to support the need for IDP modeling.

\subsubsection{Scen6: Human-Outdoor}
The measurement is conducted at an outdoor site at Beijing University of Posts and Telecommunications. The surrounding environment includes scatterers such as trees, vehicles, and buildings, which generate reflections and scattering. The Tx and Rx are fixed in the same location, both facing the target throughout the measurement. A male target with a height of 1.7 m and a shoulder width of 0.4 m is selected for the experiment. 
Fig.~\ref{figs:八个场景的测量照片}(f) illustrates the measurement setup. The initial distance between the Tx/Rx and the target is 1 m, with the maximum distance being 12 m. Measurements are taken at 0.1 m intervals between each adjacent position, resulting in a total of 111 measurements, covering various distances from near to far.

\subsubsection{Scen7: Indoor hall}
The measurement is conducted in a typical indoor hall.
Fig.~\ref{figs:八个场景的测量照片}(f) illustrates the measurement setup.
The measurements are taken under both mono-static and bi-static scenarios. In the mono-static setup, the horn antennas for both the Tx and Rx are placed at the same location in the center of the hall,  In the bi-static setup, the horn antenna at the transmitter remains at the center of the hall, while the omnidirectional antenna, acting as the receiver, is positioned at three different locations.

To extract the channel information from different directions, the horn antenna rotates horizontally in a clockwise direction. These rotations cover angles from 0° to 360°, with an increment of 5°, resulting in a total of 72 angles at which the channel is measured.

\section{Target channel modeling methods and measurement validation}
\subsection{Concatenation}
\subsubsection{Concatenation of path loss}

\begin{figure}[ht]
    \centering
    \includegraphics[width=\columnwidth]{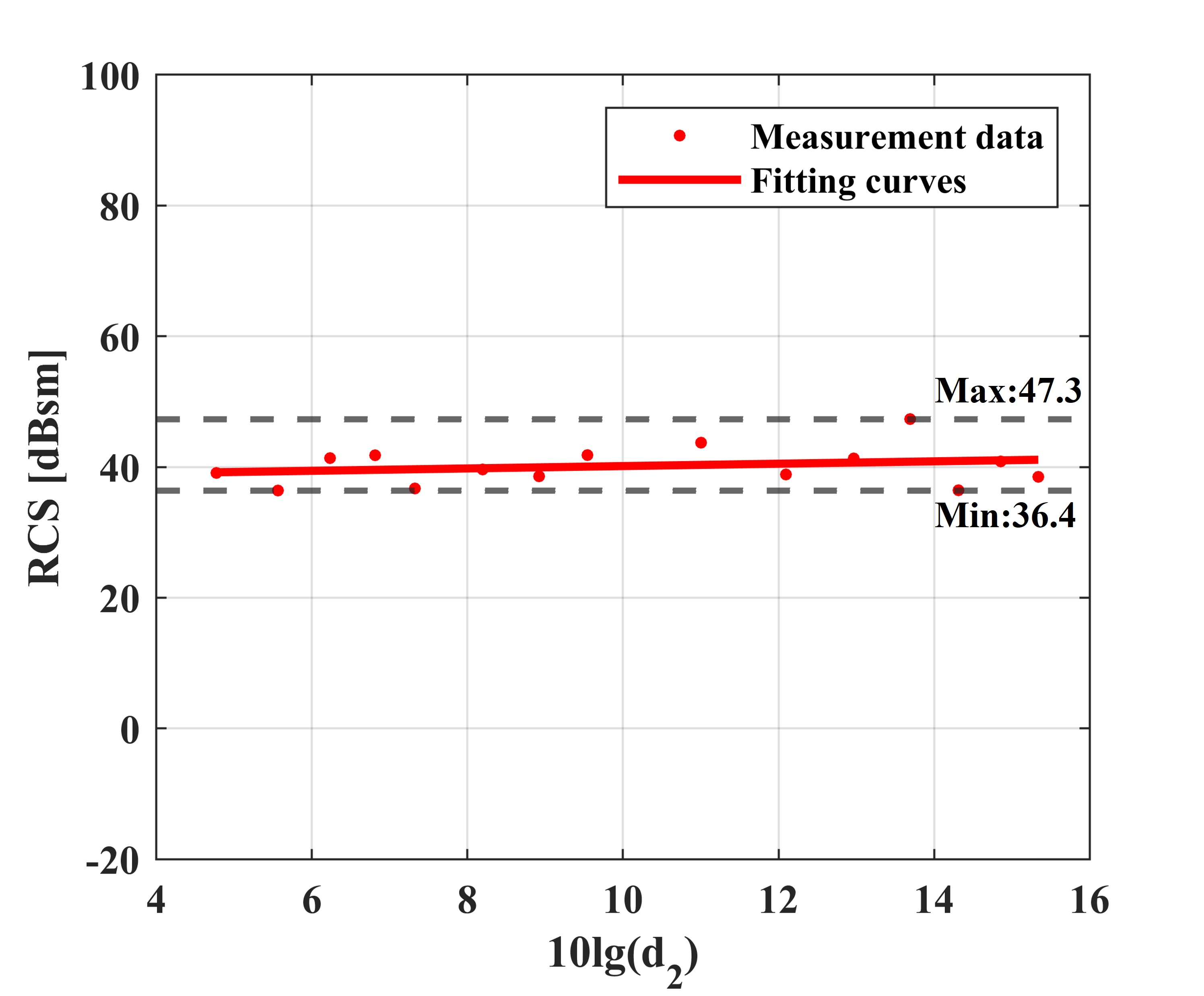}
    \caption{RCS fitting curve.}
    \label{fig:级联路损的RCS拟合直线}
\end{figure}

\addtolength{\topmargin}{0.04in}
\begin{table*}[h]
\centering
\newcommand{\tabincell}[2]{\begin{tabular}{@{}#1@{}}#2\end{tabular}}
\caption{Comparison of Results Under Codebook 2\cite{ref:张骥威RIS级联}}
\renewcommand{\arraystretch}{1.5}
\small
\tabcolsep=0.3cm
\begin{tabular}{c|c|c|c|c|c|c|c}
\hline
\hline
\textbf{Path} & \textbf{ $\bm{ P_{n_1} }$ (dB)} & \textbf{ $\bm{ P_{n_2} }$ (dB)} & \textbf{$\bm{ \sigma_\text{ST} }$ (dB)} &  \textbf{\tabincell{c}{Concatenated Power \\ $\bm{ P^{\text{conv}}_{n_1,n_2} }$ (dB)}} & \textbf{\tabincell{c}{Actual Received \\ Power $\bm{ P_{n} }$  (dB)}} & \textbf{\tabincell{c}{ Received Power \\ without RIS (dB)}} & \textbf{\tabincell{c}{Difference \\ $\bm{ \Delta P }$  (dB)}} \\ \hline
1-A & -74.64 & -78.46 & 8.48 & -106.39 & -107.54 & -135.38 & 1.15 \\ \hline
2-A & -70.21 & -78.46 & 9.04 & -101.40 & -106.14 & -150.48 & 4.74 \\ \hline
1-B & -74.64 & -83.36 & 14.19 & -105.58 & -112.18 & -151.06 & 6.60 \\ \hline
2-B & -70.21 & -83.36 & 4.46 & -110.88 & -105.81 & -142.97 & -5.07 \\ \hline
1-C & -74.64 & -93.28 & 0.46 & -128.94 & -128.83 & -141.26 & -0.11 \\ \hline
2-C & -70.21 & -93.28 & -6.33 & -118.56 & -113.47 & -147.06 & -5.09 \\ \hline
1-D & -74.64 & -95.59 & 0.75 & -131.54 & -132.19 & -143.43 & 0.65 \\ \hline
2-D & -70.21 & -95.59 & 6.70 & -133.90 & -134.53 & -139.08 & 0.63 \\ \hline
\hline
\end{tabular}
\label{compare}
\end{table*}

\label{asec:Co}
When modeling the path loss of the target channel, the classical radar equation can be used, which is expressed as follows:
\begin{equation}
PL^{\textnormal{tar}} = PL^{\textnormal{Tx-ST}}(d_1) + PL^{\textnormal{ST-Rx}}(d_2) + 10 \log_{10} \left( \frac{\lambda^2}{4 \pi} \right) - \sigma_{\textnormal{RCS}},
\label{eq:雷达方程}
\end{equation}
where $d_1$ and $d_2$ represent the distances between Tx and the target, and between the target and Rx. $PL^{\textnormal{tar}}$, $PL^{\textnormal{Tx-ST}}(d_1)$ and $PL^{\textnormal{ST-Rx}}(d_2)$ are the path losses of the Tx-ST-Rx, Tx-ST, and ST-Rx links, respectively.

To verify the applicability of the radar equation in the communication domain, equation~\eqref{eq:雷达方程} is reformulated as follows:
\begin{equation}
\sigma = PL^{\textnormal{Tx-ST}}(d_1) + PL^{\textnormal{ST-Rx}}(d_2) - PL^{\textnormal{tar}} + 10 \log_{10} \left( \frac{\lambda^2}{4 \pi} \right).
\label{eq:雷达方程变形}   
\end{equation}

Since the target remains unchanged during the measurement, it is assumed that its RCS is constant. The key to verifying the radar equation lies in whether the right-hand side of equation~\eqref{eq:雷达方程变形} remains constant as the distance varies. The path loss of the target channel and its sub-links (Tx-ST and ST-Rx) has been obtained through channel measurements. Taking the measurement result where the distance between the Tx and the target (metal plate) is 4 m as an example, Fig.~\ref{fig:级联路损的RCS拟合直线} shows the RCS curve of the target as a function of $d_2$. Theoretically, the RCS of the target should be a horizontal line, independent of $d_2$, i.e., a straight line with zero slope. However, due to the influence of the waveguide-like structure of the corridor\cite{ref:李毅波导}, the data points at different distances exhibit a certain degree of fluctuation., with a minimum value of 36.4 dBsm and a maximum value of 47.3 dBsm. Despite this, the fitting result shows that the overall trend of the data is close to a straight line with zero slope. Therefore, it can be concluded that the radar equation is also applicable in the target channel path loss modeling, reflecting the concatenation relationship between the Tx-ST and ST-Rx links and the influence of the target's RCS.

\subsubsection{Convolution of multipath fast fading}
In the experimental validation, the antenna gain is already removed, and the path power can be expressed as
\begin{align}
P_{n}=|h_{\text{tar}}(\tau_{n})|^2,
\end{align}
where \( h_{\text{target}} \) represents the measured target channel CIR, \( P_{n} \), \( \tau_{n} \) represent the power and delay of \( n^{th} \) path in target channel, respectively.

To assess whether the concatenation channel model can effectively describe reality in scenarios involving multi-hop, we conduct a channel measurement campaign in a factory scenario with the target of a RIS in order to change the target RCS. The setup of the target channel measurement system and details of the scenario are as described before. Since the analysis is performed at the path level, the path subscript can be simplified from $(n_1,m_1 )^{th}$  to $n_1^{th}$.

The sub-link path power \( P_{n_1} \) and \( P_{n_2} \) can be obtained from the measured sub-link CIR in a similar manner.

The theoretical concatenated path power can be calculated as
\begin{align}
P^{\text{conv}}_{n_1,n_2}=P_{n_1}+P_{n_2}+\sigma_{\text{RCS}}(\theta_{n_2}^{\text{out}}, \theta_{n_1}^{\text{in}})-10 \log_{10} \left( \frac{\lambda^2}{4\pi} \right),
\label{eq:级联径功率计算}
\end{align}
where $P^{\text{conv}}_{n_1,n_2}$  refers to the power calculated through the  concatenation channel model, and all terms are expressed in the form of power or power gain in dB.

The key to validating the accuracy of the concatenated model lies in analyzing the difference between the theoretical path power and the measured path power.
\begin{align}
\Delta P=P^{\text{conv}}_{n_1,n_2}-P_{n},
\end{align}
If this difference falls within an acceptable range, the model can be considered to fit the measured data well.

During the rotational measurements, the power delay profile (PDP) at specific angles can be expressed as
\begin{align}
PDP_{\theta}(\tau)=|h_{\theta}(\tau)|^2,
\end{align}
where $h_{\theta_0}(\tau)$ represents the CIR measured at the corresponding angle.

To further visualize the PDP at different angles, the power-angle delay profile (PADP) is expressed as
\begin{align}
PADP(\theta,\tau)=\sum\delta(\theta-\theta_n)\cdot PDP_{\theta_n}(\tau).
\end{align}

\begin{figure}[!htbp]
    \centering
    \subfloat[]{\includegraphics[width=0.5\columnwidth]{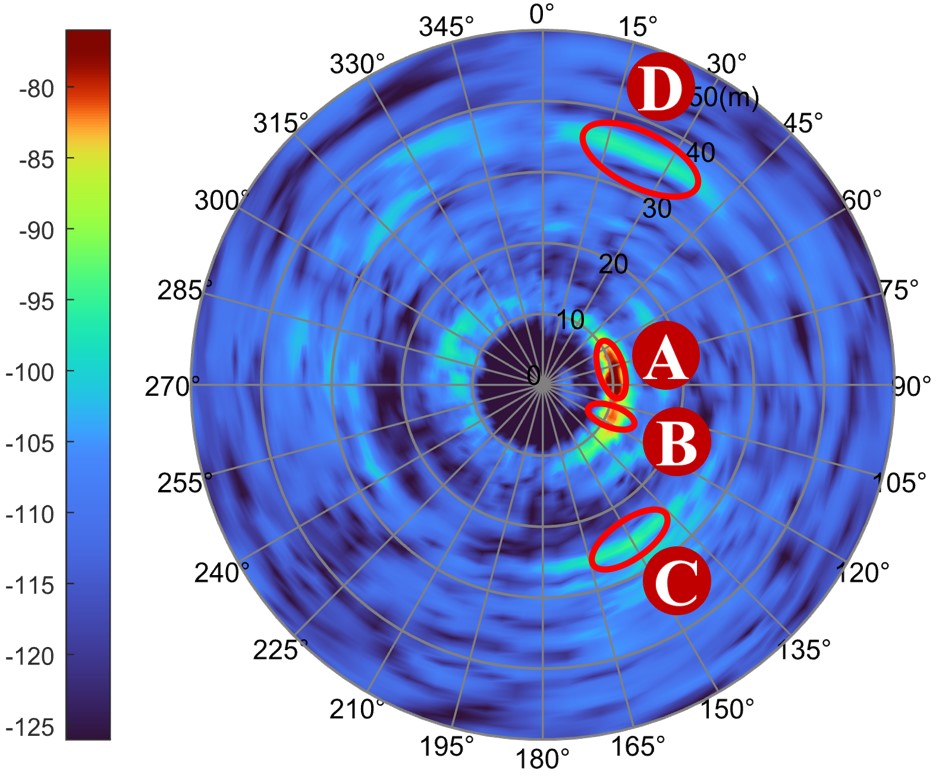}}
    \hfill
    \subfloat[]{\includegraphics[width=0.5\columnwidth]{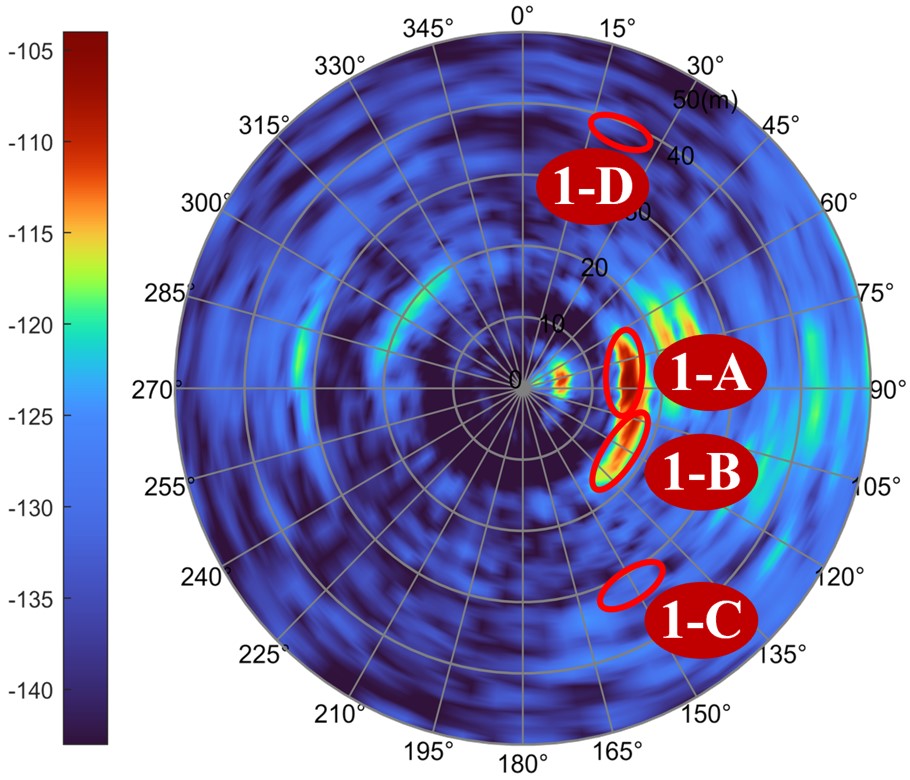}}
    \caption{PADPs of (a) ST-Rx measurement, (b) Tx-ST-Rx measurement\cite{ref:张骥威RIS级联}.}
    \label{figs:小尺度级联验证PADP}
\end{figure}

Figure~\ref{figs:小尺度级联验证PADP}(a) shows the power angle delay profile (PADP) results of the ST-Rx subchannel, where four distinct multipath signals are visible, labeled as Path A to Path D. Through path reconstruction analysis, Path A is identified as the LOS path, while Paths B, C, and D are reflection paths caused by large environmental scatterers. In comparison, Figure~\ref{figs:小尺度级联验证PADP}(b) presents the PADP results for the overall target channel when the transmitter is located at position 1. Similar multipath signals can be observed at the corresponding angles of arrival for Paths A to D, labeled as Paths 1-A to 1-D. Although the presence of Paths 1-C and 1-D is less prominent in the PADP, a comparison with the environmental PADP reveals that the power of these two paths increases by approximately 10 dB relative to the background channel, indicating that the RIS indeed introduces multipath signals in these directions.

To verify whether Paths 1-A to 1-D are the result of cascading the LOS path in the Tx-ST segment with Paths A to D, an initial delay analysis is performed. It is found that the delays of the four paths in the target channel are extended by 17.5 ns compared to the corresponding paths in the ST-Rx subchannel, which matches the delay of the LOS path in the Tx-ST segment. This provides indirect evidence supporting the concatenation path hypothesis.

For further quantitative analysis, the RCS of the RIS is calculated using the physical optics method from\cite{ref:周正福RIS}, and the differences between the theoretical concatenated power and the measured power are compared. By comparing the calculated concatenated power with the actual measured power, it is observed that the difference does not exceed 7dB, with the smallest being only 0.11dB. Considering an angular measurement interval of 5°, which leads to a maximum angular error of 2.5° at \(\theta^\text{out}_{n_2}\), and given the significant fluctuation range of the RCS of RIS, a certain level of difference is considered acceptable.
The detailed comparison results are shown in Table 37. The table also lists the received power measured at the corresponding angles and delays for each path in the absence of the RIS. By comparing the results with and without the RIS, it is clear that the RIS introduces concatenated paths in multiple directions corresponding to Paths 1-A to 2-D. This one-to-one path matching characteristic is a significant feature of the concatenated convolution model.

\subsection{Direct and indirect path}
\begin{figure*}[t]
	\centering
	\subfloat[Human-Indoor Pos 3]{\includegraphics[width=0.33\textwidth]{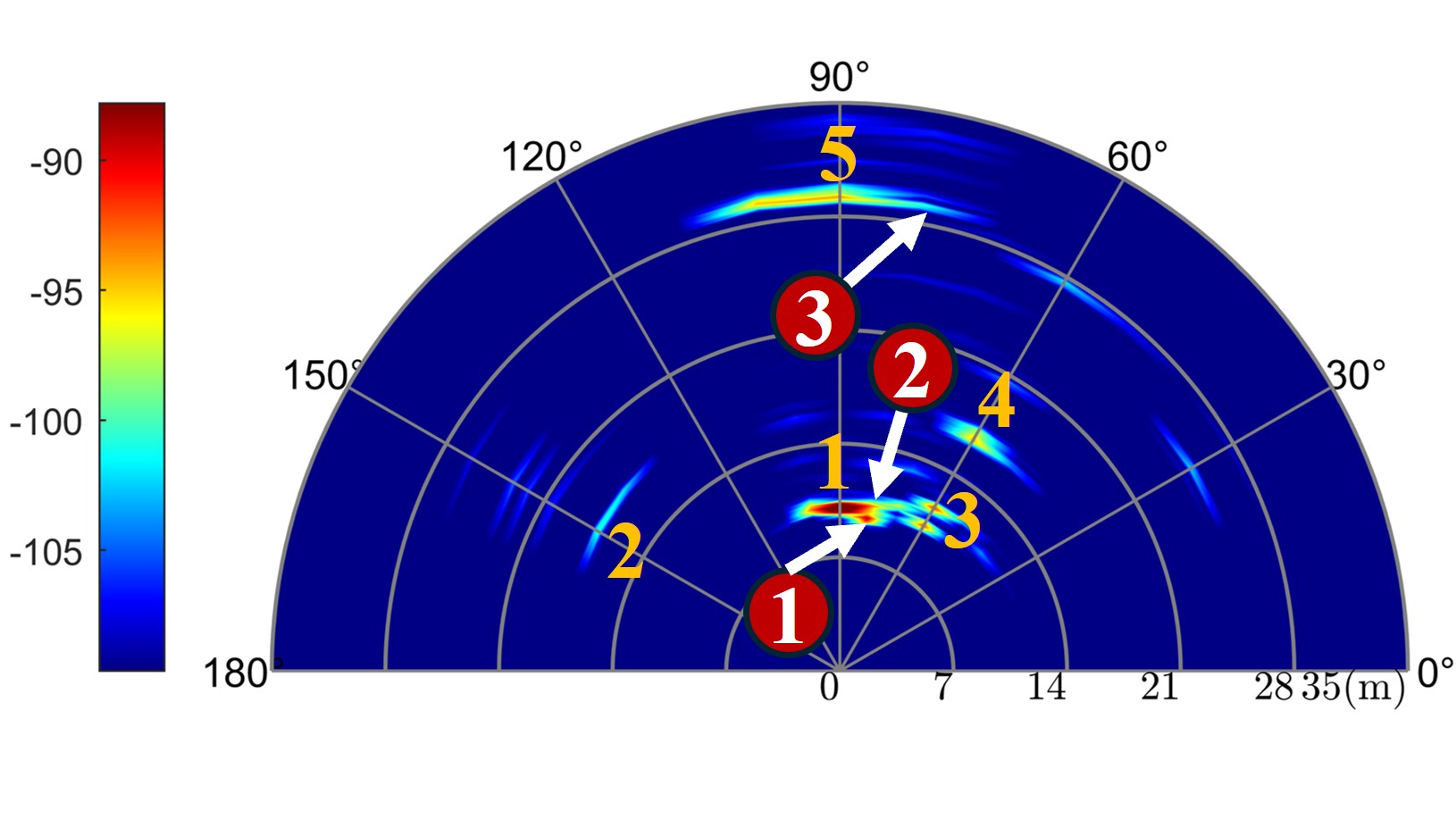}}
	\hfill
	\subfloat[Human-Indoor Pos 13]{\includegraphics[width=0.33\textwidth]{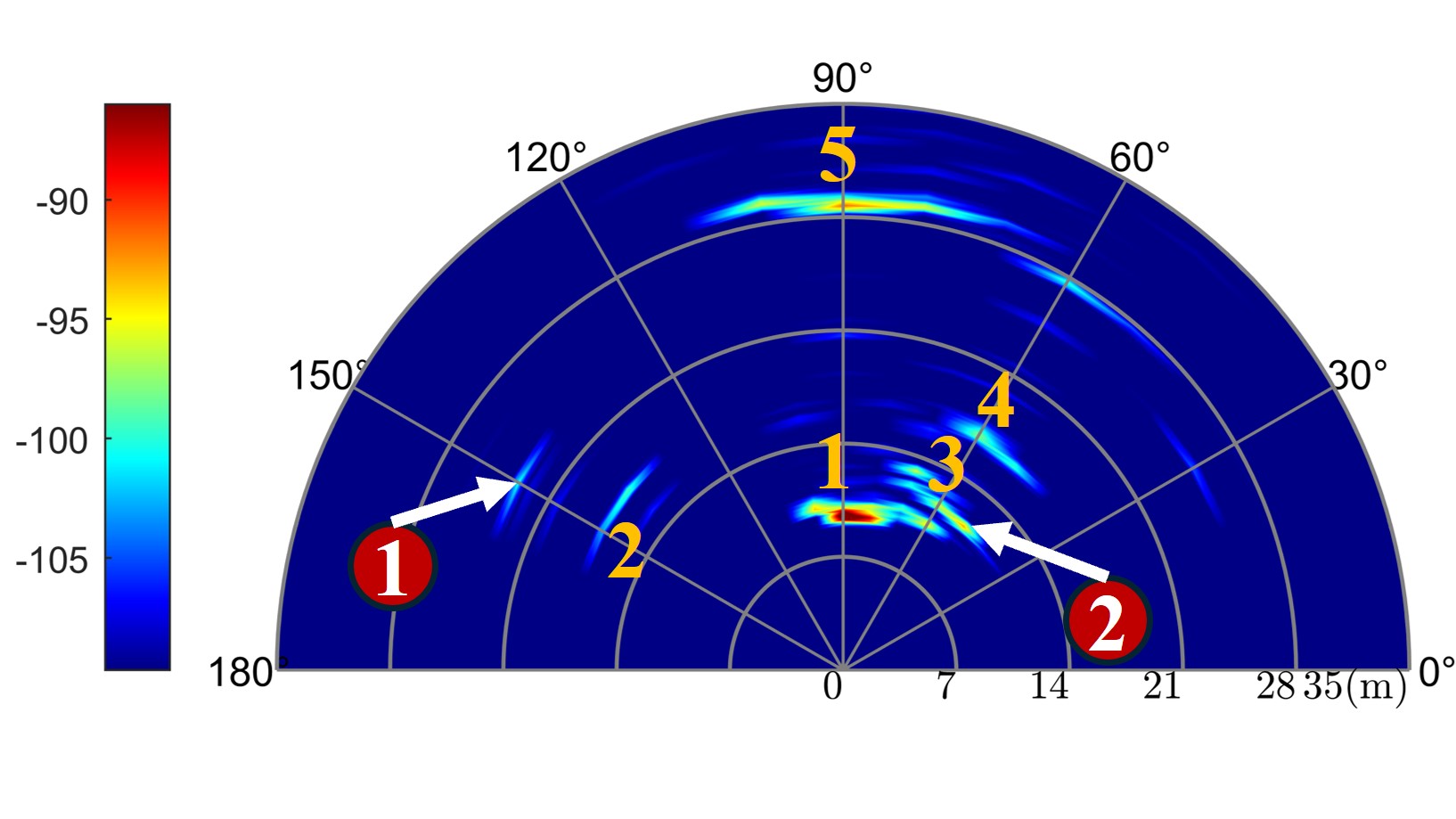}}
	\hfill
	\subfloat[UAV-UMi mono-static]{\includegraphics[width=0.33\textwidth]{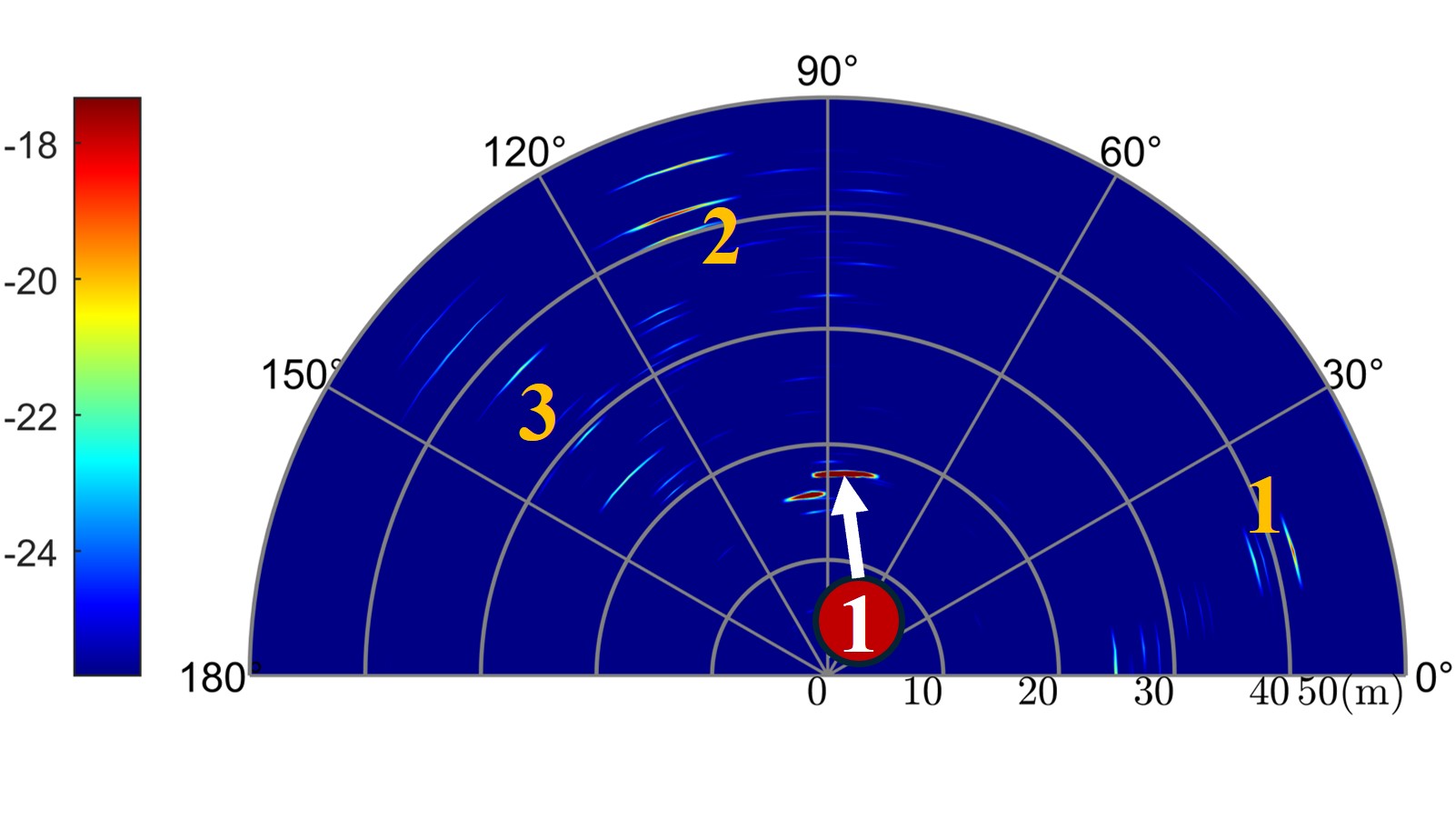}} \\[0.3cm] 
	\subfloat[UAV-UMi bi-static]{\includegraphics[width=0.33\textwidth]{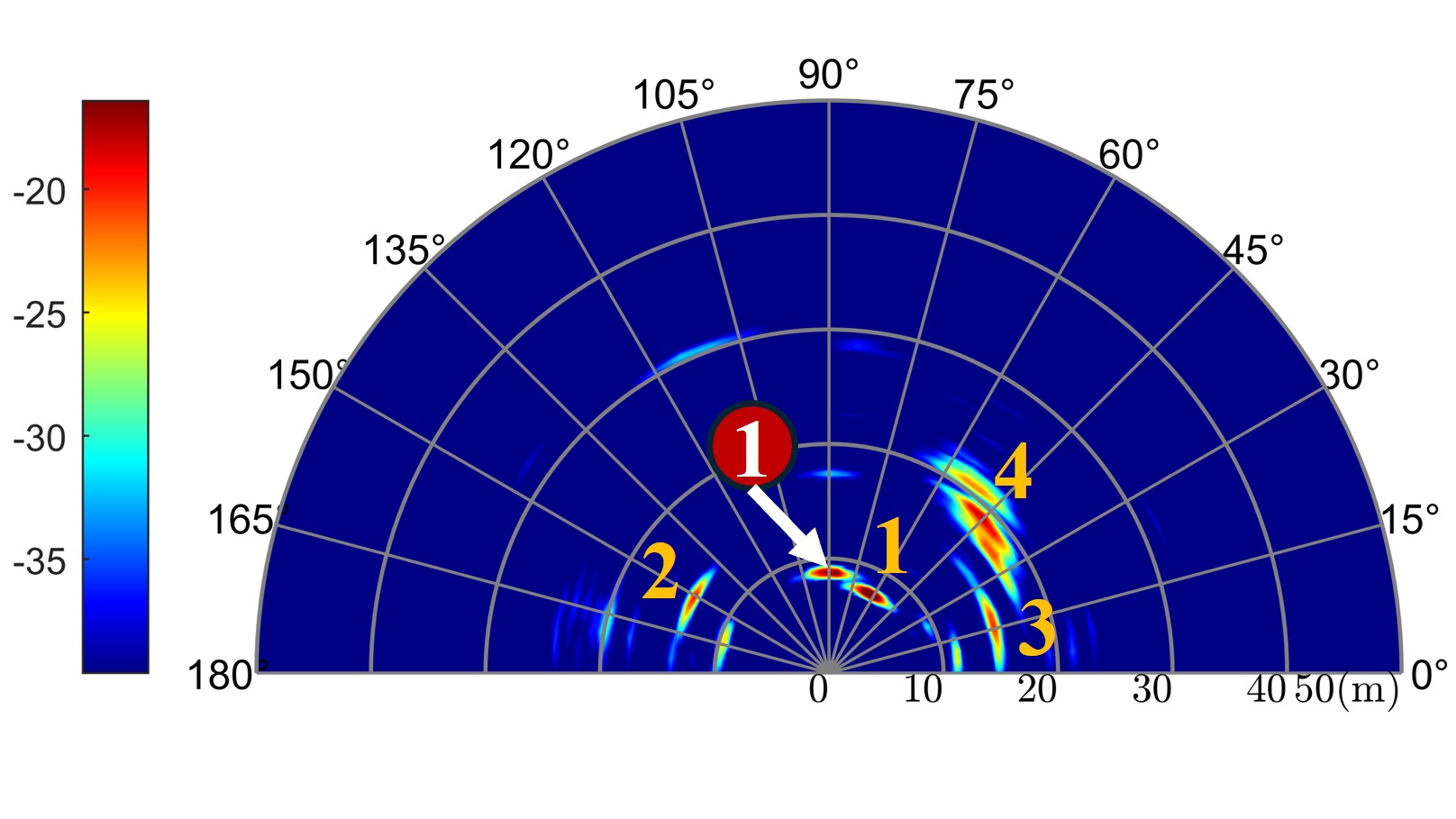}}
	\hfill
	\subfloat[Vehicle-Outdoor case 1]{\includegraphics[width=0.33\textwidth]{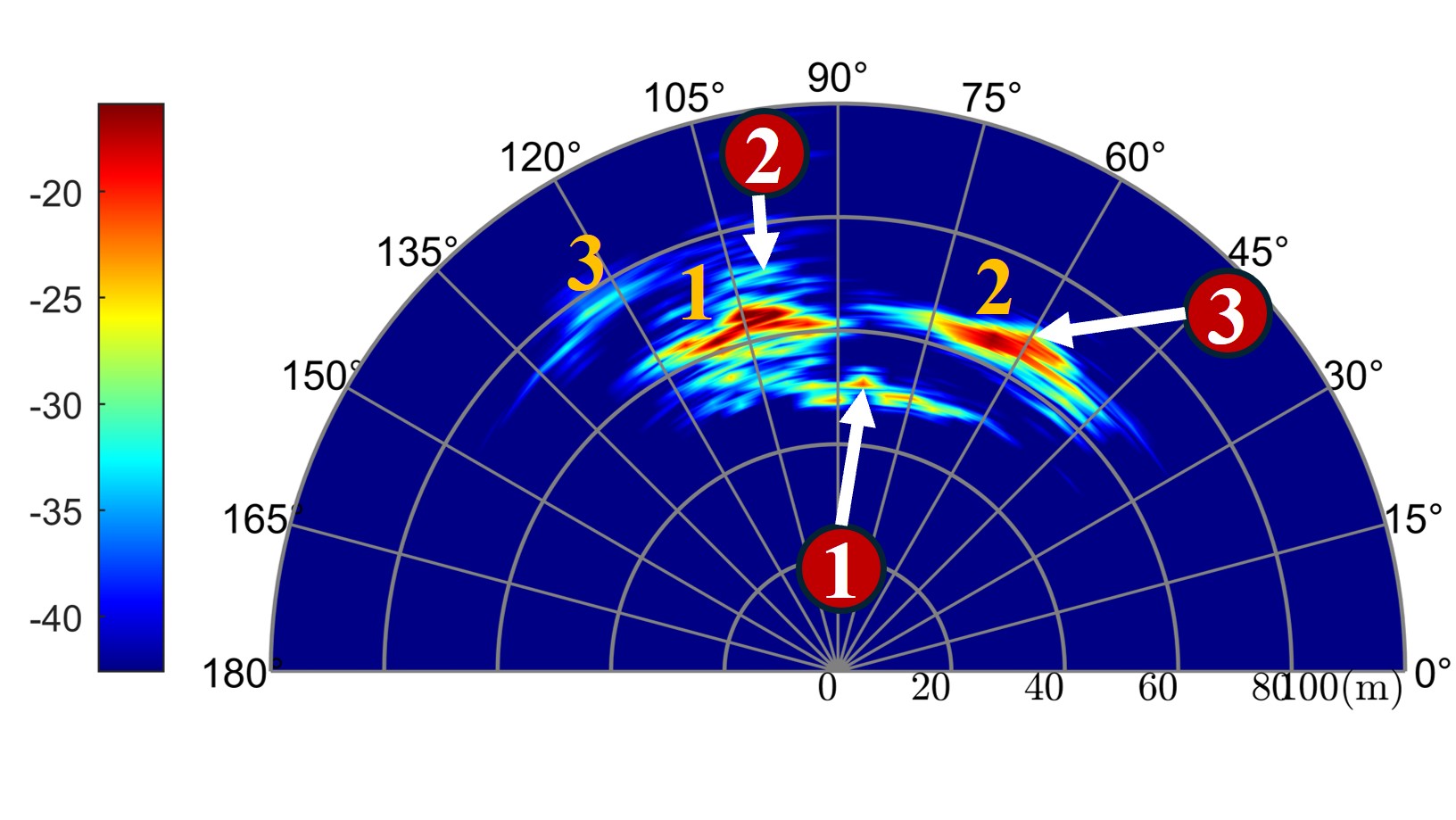}}
	\hfill
	\subfloat[Vehicle-Outdoor case 2]{\includegraphics[width=0.33\textwidth]{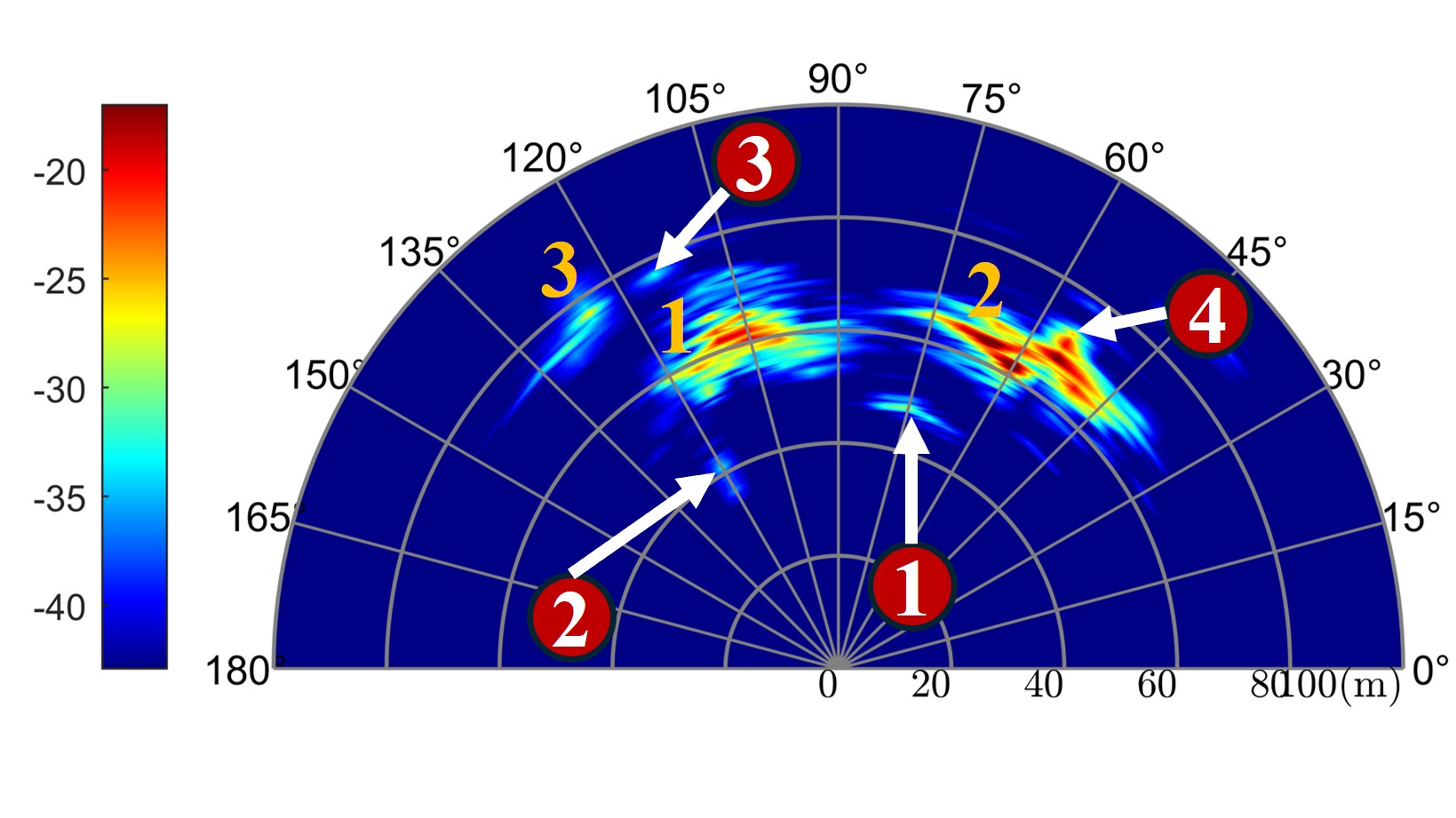}}
	
	\caption{The PADPs of sensing channels in three typical scenarios. The numbers marked with red circles indicate paths belonging to the target channel, while the numbers without red circles indicate paths belonging to the background channel.}
	\label{figs:四种场景八张PADP}
\end{figure*}

To investigate the importance of modeling IDPs in the target channel, we conducted field measurements in three typical sensing scenarios: Human-Indoor, UAV-UMi and Vehicle-Outdoor. First, we extracted paths that belonged to the target channel by comparing them to background channels, ensuring precise identification. The measurements are visualized using PADP. Then, we reconstructed the signal propagation paths based on geometric optics theory, differentiating between direct and indirect paths. Finally, we evaluated the power ratio between direct and indirect paths.

Considering the modeling of IDPs, the target channel can be represented as the superposition of DPs and IDPs:
\begin{align}
h_{u,s}^{\textnormal{tar}}(t,\tau) = \sqrt{P_0}\delta(\tau - \tau_0) + \sum\sqrt{P_i}\delta(\tau - \tau_i),i=1,2,3...,
\end{align}
The parameters $P_0$ and $\tau_0$ represent the power and delay of the direct path, respectively, while $P_i$ and $\tau_i$ represent the power and delay of the $i$-th order non-direct path, where $i$ indicates the number of interactions the signal undergoes with objects other than the target during propagation.
The power ratio of Dps and IDPs of different orders are defined as follows:

\begin{align}
PP_i = \frac{P_i}{P_{\text{tar}}}, \quad i=0,1,2,\dots,
\label{eq:功率占比计算方法}
\end{align}

Here, $P_{\text{tar}}$ represents the total power of the target channel, and $PP_i$ represents the proportion of the direct path and each order of non-direct paths. In the subsequent analysis, the focus was primarily on the direct path ($i=0$), first-order IDPs ($i=1$), and other higher-order IDPs ($i\geq2$).

\subsubsection{PADP analysis}
Fig.~\ref{figs:四种场景八张PADP} presents the PADP of the ISAC channels in three typical scenarios. The results include the path information from both the target and background channels. For each scenario, two representative cases are selected for detailed analysis. By comparing the PADP of each case with that of the background channel, the multipath components in the target channel are accurately extracted. In PADP, multipaths belonging to the background channel are labeled with numbers, while multipaths belonging to the target channel are labeled with numbers enclosed in circles. A detailed analysis of the PADP visualization results for each scenario is provided below.

In the Human-Indoor scenario, Fig.~\ref{figs:四种场景八张PADP}(a) shows the PADP results at position 3, where there is a LOS path between the Tx and the target, as well as between the target and the Rx. Fig.~\ref{figs:四种场景八张PADP}(b) presents the PADP results at position 13, where there is an LOS path between the Tx and the target, but a NLOS path between the target and the Rx. It can be observed that, in this case, no obvious DP is observed, and all the paths in the target channel can be classified as IDP.

In contrast, for the UAV-UMi scenario, Fig.~\ref{figs:四种场景八张PADP}(c) shows the results for a mono-static setup, and Fig.~\ref{figs:四种场景八张PADP}(d) presents the results for a bi-static setup. It is evident that the target channel primarily exhibits LOS paths, or DPs.

For the Vehicle-Outdoor scenario, Fig.~\ref{figs:四种场景八张PADP}(e) corresponds to case 1, where the transmitting and receiving antennas are both directly facing the target. The target is positioned at a 45° angle at the corner of the intersection between the Tx and Rx, creating a mirror-like reflection for the LOS path between the antennas and the target. Fig.~\ref{figs:四种场景八张PADP}(f) illustrates the scenario where the target is moved or its orientation is changed, resulting in changes to the multipath components in the target channel. In this case, DP no longer dominates absolutely, and through non-mirror scattering from the vehicle body, more IDPs are generated.

\subsubsection{Paths reconstruction}
This section analyzes the reconstructed multipath propagation paths in the target channel based on the PADP visualization results. Following the classification method described earlier, the multipaths are divided into three categories: the DPs ($i=0$), first-order IDPs ($i=1$), and other higher-order IDPs ($i\geq2$). A detailed analysis of the power contribution of each type of path is provided. Next, the method for reconstructing propagation paths will be introduced using the Human-Indoor scenario as an example.

\begin{figure}[ht]
    \centering
    \includegraphics[width=\columnwidth]{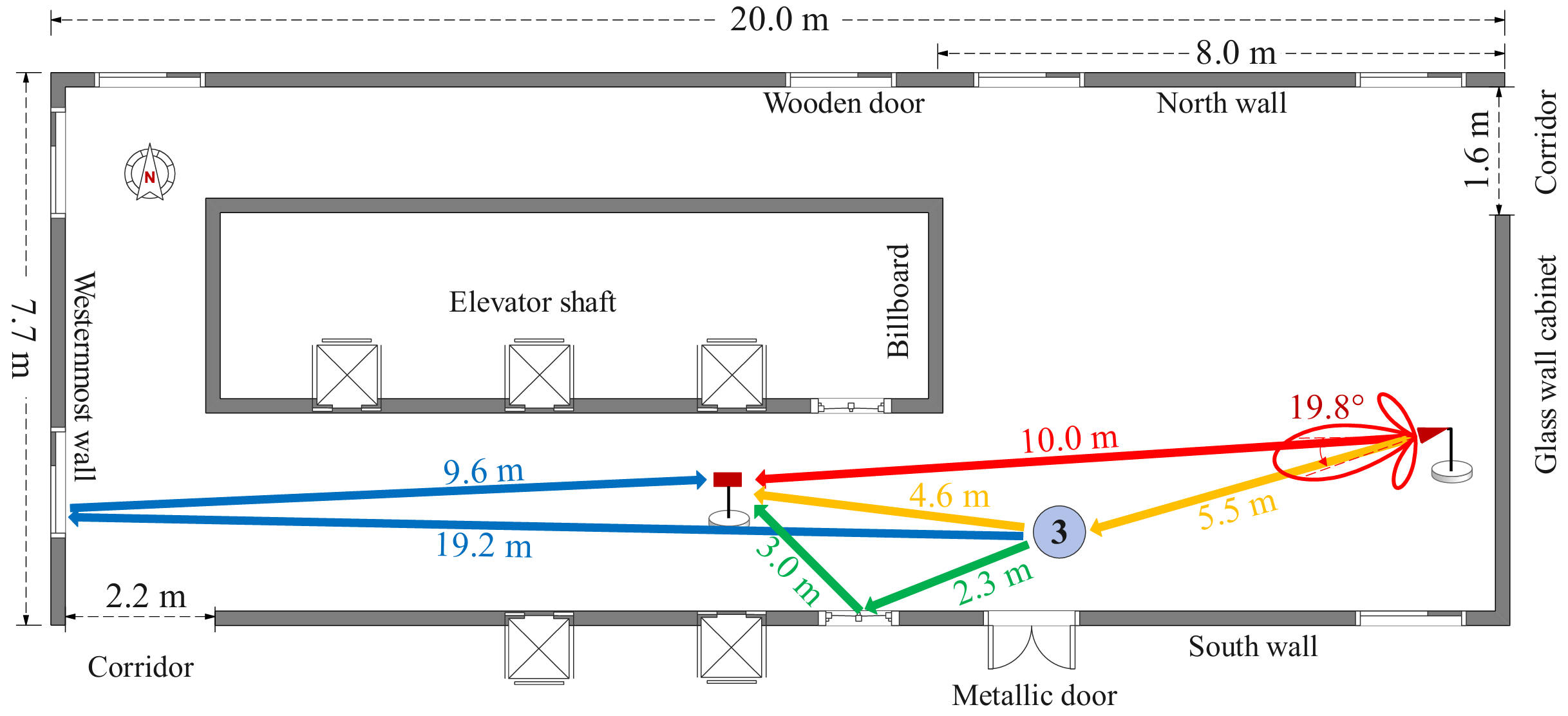}
    \caption{Illustration of the potential propagation paths at Pos 3. The lines with different colors represent propagation paths of multipath components (MPCs)\cite{ref:陈文俊人体}.}
    \label{fig:室内人体路径追踪}
\end{figure}

\textit{Human-Indoor}
In Pos 3, as shown in Fig.~\ref{figs:四种场景八张PADP}(a) The AoD for Path \circled{1} is 80°, and the delay is 33.3 ns, corresponding to a propagation distance of 10 m. Based on geometric optics theory, the propagation route of Path \circled{1} can be reconstructed using this delay. Given that the horn antenna used in the experiment has a 3dB beamwidth of 19.8°, and by tracing back through the scene diagram, two propagation paths that match the angle and delay conditions can be identified in Fig.~\ref{fig:室内人体路径追踪}, marked by red and orange lines respectively. The red path represents the Tx→Rx background channel path with a propagation distance of 10.0 m. The orange path is Tx→target→Rx with a propagation distance of 10.1 m. The delay difference between these two propagation paths is smaller than the delay resolution of the experimental measurements, so they are determined to be a single path. Similarly, Paths \circled{2} corresponds to the Tx→target→south wall→Rx green path, with a propagation distance of 10.8 m. Paths \circled{3} corresponds to the Tx→target→west wall→Rx blue path, with a propagation distance of 34.3 m.

To verify the necessity of modeling NLOS paths in the target channel, the power ratio of the LOS path and various orders of NLOS paths is calculated according to equation~\eqref{eq:功率占比计算方法}. Furthermore, a similar path reconstruction analysis is performed at position 13. The power ratio distribution of each order of paths at the two positions is shown in the human target part of Table~\ref{tab:四种目标非直射径功率对比}. At position 13, where the target and Rx are in a NLOS condition, no LOS path exists, and the power ratio of higher-order NLOS paths increases significantly.

Path reconstruction analysis confirms that the interaction between the target and environmental scatters leads to the formation of IDPs in the target channel. In the case of human targets, reflected paths play a significant role in the target channel. For example, in position 3, IDPs account for over 80\% of the total power, while at position 13 (which includes both LOS and NLOS paths), all received signals come from IDPs. This phenomenon indicates that IDPs play an important role in the target channel.

Therefore, target channel modeling for human sensing must consider the presence of IDPs, especially in the ST-Rx link. Relying solely on the LOS path model is insufficient to accurately describe the propagation characteristics of the target channel, and modeling NLOS paths is key to improving the accuracy of channel modeling.

For vehicle and UAV measurements, we applied the same path reconstruction method to analyze the NLOS paths in different target channels.

\begin{table*}[ht]
	\caption{Power proportion of each bounce MPCs of target-Rx link}
	\newcommand{\tabincell}[2]{\begin{tabular}{@{}#1@{}}#2\end{tabular}}
	\centering
	\renewcommand{\arraystretch}{1.5}
	\small
		\tabcolsep=0.3cm
		\begin{tabular}{c|c|c|c|c|c|c}
			\hline 
			\hline
			\textbf{Target Type} & \multicolumn{2}{c|}{\textbf{Human}} & \multicolumn{2}{c|}{\textbf{UAV}} & \multicolumn{2}{c}{\textbf{Vehicle}} \\ \hline
			Case & \tabincell{c}{Position 3 \\ (LOS+LOS)} & \tabincell{c}{Position 13 \\ (LOS+NLOS)} & Mono-static & Bi-static & Case 1 & Case 2 \\ \hline
			$PP_0$          & 18.9\%    & 0\%    & 100\%    & 100\%    & 50.5\%    & 17.9\%    \\ \hline
			$PP_1$          & 65.7\%    & 16.4\%    & 0\%    & 0\%    & 15\%    & 8.4\%     \\ \hline
			$PP_{i \ge 2}$  & 15.4\%    & 83.6\%    & 0\%    & 0\%    & 34.5\%    & 73.7\%    \\ \hline
			\hline
		\end{tabular}
	\label{tab:四种目标非直射径功率对比}
\end{table*}

\textit{UAV-UMi}
During the UAV-UMi mono-static sensing measurements, the Rx and Tx are located at the same position, receiving echo signals reflected by the UAV and the environment. By comparing the PADP in Fig.~\ref{figs:四种场景八张PADP} (c) for scenarios with and without the UAV, it can be observed that path \circled{1} corresponds to the DP of the target channel, while paths 1, 2, and 3 belong to the background channel. Notably, the IDPs in the UAV target channel are not significant, and the UAV has little impact on the background channel. When the UAV is at a height either lower or higher than the Tx/Rx (antenna vertical angles of -15° and 15°), IDPs are barely noticeable, with the target’s DP only visible at a 90° angle.

In the bi-static sensing mode, the UAV hovers between the sensing Tx and Rx, at distances of 17.4 m from Tx and 17 m from Rx. Path \circled{1} is the direct path in the UAV target channel, corresponding to the Tx→ST→Rx route. Additionally, path 1 represents the Tx→Rx background path, which has a higher power. Paths 2, 3 and 4 are multiple reflection paths in the background channel. In this set of bi-static sensing measurements, no significant IDPs are observed either.

Based on outdoor ISAC channel measurements targeting UAVs, clear DPs are observed in both mono-static and bi-static sensing modes, while the influence of IDPs is not significant. Therefore, in channel modeling for UAV scenarios, the focus should primarily be on the DPs. Further measurements are needed to explore the modeling of IDPs in UAV scenarios.

\textit{Vehicle-Outdoor.}
In the vehicle scenario, both the Tx-ST and ST-Rx links are in LOS conditions in case 1, while in case 2, the Tx-ST link remains LOS, but the ST-Rx link becomes NLOS. Through path reconstruction analysis, the power ratio of the DPs and various orders of IDPs is obtained, and the relevant results are summarized in the vehicle scenario part of Table~\ref{tab:四种目标非直射径功率对比}.

In case 1, it might be intuitively expected that the direct Tx-ST-Rx path would dominate. However, experiments proved otherwise. Due to the metallic and highly smooth surface of the vehicle, the signal reflection exhibits a mirror-like effect, heavily dependent on the incident angle. Additionally, with a distance of about 22 m between the transmitter and the vehicle, the signal beamwidth at the vehicle is approximately 4 m. Part of the signal is directly reflected from the vehicle surface to the receiver, while other part is reflected off the sides or top of the vehicle, eventually reflecting off a distant fence. For example, Path 2 is formed by reflections from a lamppost and a wall behind the vehicle, consisting of both first-order and second-order IDP; Path 3 is the signal reflected from the wall behind the vehicle, primarily considered as a first-order IDP. Therefore, although the vehicle blocks some multipath components that would have propagated directly from the transmitter to the distant fence, the reflections from the vehicle actually increase the number of multipath components forming the sensing channel between the vehicle and the fence, with the majority of the channel power coming from these reflections.

In case 2, since the Tx antenna is not aligned with the vehicle, part of the signal propagates directly to the distant fence. The power variations caused by the fence are relatively small compared to the surrounding environment, so its power contribution is not significant. However, given the beamwidth, some signals reflect off the side of the vehicle and reach the receiver directly, or reach the receiver after being reflected by other scatterers. Path 2 is a second-order IDP from the lamppost; Path 3 is a first-order IDP from the fence; and Path 4 is a second-order IDP from the building wall. Calculations and statistical results show that a significant portion of the power comes from the signal propagating through the vehicle to the fence and then reflecting off a wall before finally reaching the receiver.

Based on the above measurement results, it can be concluded that when a vehicle is introduced as a target into the environment, existing scatterers in the environment have varying degrees of impact on the target channel. Specifically, lampposts, due to their relatively small size, contribute only about 12\% of the power in case 2 and about 5\% in case 1. In contrast, large scatterers such as walls and metal fences have a more significant influence on the target channel. In case 1, the fence contributes about 50\% of the target channel power, while in case 2, the power contribution from the wall reaches about 70\%.

Based on the ISAC channel measurements in outdoor vehicle scenarios, the presence of IDPs can be clearly observed, with the majority of the propagation power in the target channel concentrated on large scatterers like walls and fences. Therefore, in channel modeling for outdoor vehicle scenarios, priority should be given to considering the effects of both DPs and IDPs, particularly by including scatterers such as walls and fences in the target channel model.

\section{Background channel modeling methods and measurement validation}
\subsection{Background channel characteristics in mono-static sensing mode}
\begin{figure}[ht]
    \centering
    \includegraphics[width=\columnwidth]{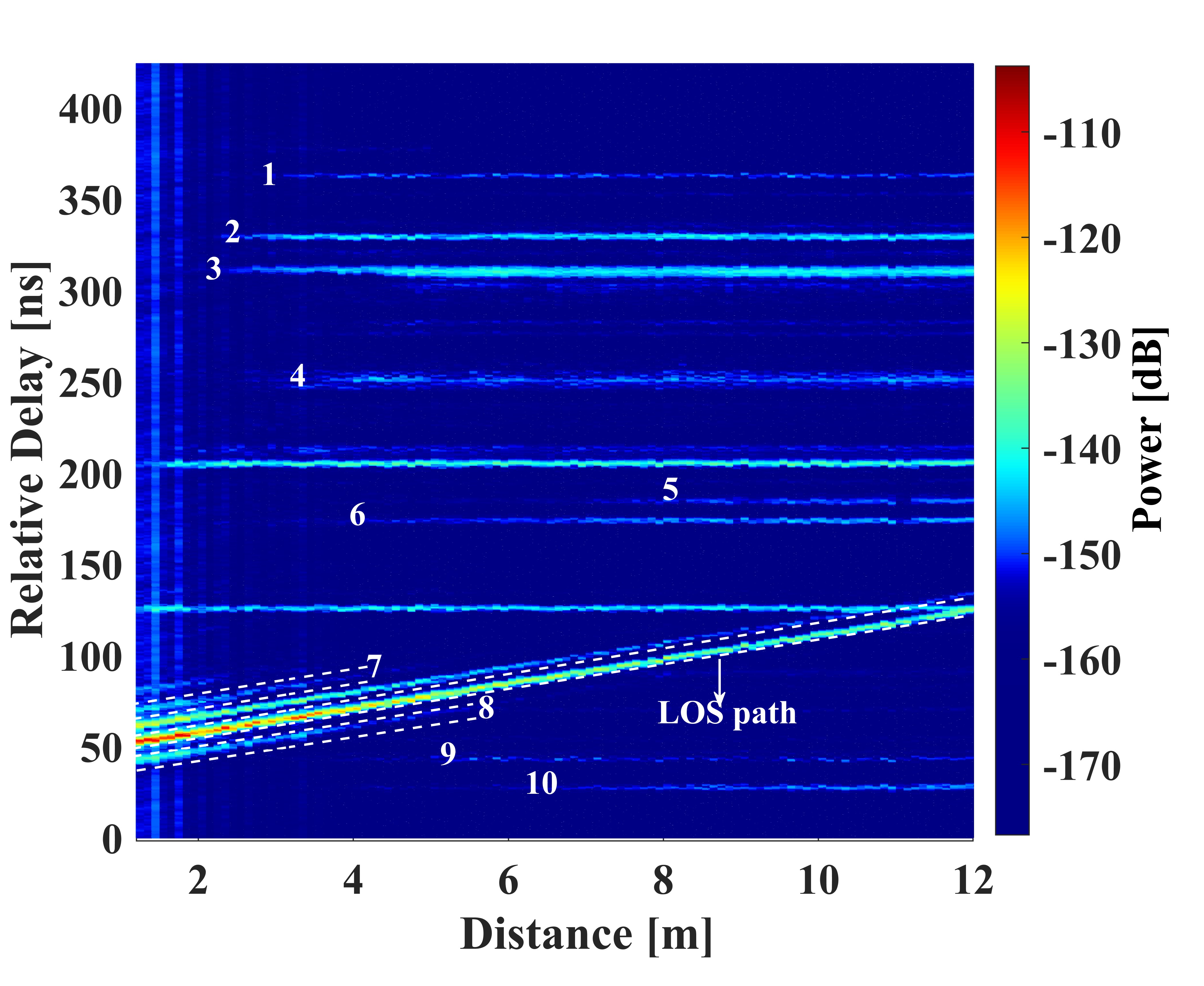}
    \caption{Comparison plot of the time delay with distance obtained from actual measurements and geometric calculations\cite{ref:王嘉琳测量}.}
    \label{fig:不同距离下PDP图}
\end{figure}

In Scenario 6: Outdoor, the PDP results at various distances are shown in Figure~\ref{fig:不同距离下PDP图}. The horizontal axis shows the distance between the human target and the Tx/Rx, while the color denotes the strength of each path. It should be noted that environmental factors are the main cause of power variances, which results in notable changes in power distribution between various sites. As the distance grows, we can see that the power of the LOS path that the ST forms drastically drops. Furthermore, specific paths may emerge or vanish at various points as the distance varies; these paths are identified by numerical tags.

Certain paths, such as those path 7 and 8, can result from multiple reflections from scatterers surrounding the target, depending on the specifics of the measurement scenario. As the distance increases, the impact of these multiple reflections diminishes, causing the corresponding paths to eventually disappear.
On the other hand, certain paths might appear as the distance increases, perhaps coming from environmental scatterers that are farther away from the target. The antennas can detect more background scatterers when the target gets farther away because it blocks the antennas less.

We may conclude from this analysis and the experimental findings that the target's blockage of background scatterers and the declining effect of multiple reflections cause the background channel to stop undergoing notable changes after a given distance. This distance is roughly 8 m in this experiment. In these situations, conventional statistical channel modeling techniques can be used to represent the background channel. However, the mono-static background channel is affected by target-related effects such blockage and dispersion as the target moves at closer ranges. Further investigation into the relationship between the background channel and the target in mono-static sensing settings is necessary since the background channel shows fluctuations that correspond with the target's distance.

\subsection{Coupling properties between target and background channels}
In Scen3: Human-InH, based on the study of the multipath characteristics of the bi-static target channel, it was further found that the multipath of the bi-static background channel is also, to some extent, influenced by target-related effects such as scattering and blockage, exhibiting characteristics related to the target's position. This correlation is manifested at the large scale as a correlation in path loss, which can be reflected by the $O_{\text{back}}$ parameter in equation~\eqref{eq:O因子描述的路损模型}.
The effect of human targets on background channel path loss is analyzed and quantified. By comparing the PADP under conditions with and without a human target, it is observed that the presence of a human significantly alters the quantity and power of MPCs of the background channel, thus impacting the channel’s path loss characteristics. Table~\ref{tab:O因子具体数值} shows the PCF values measured under various positions, revealing a clear difference in PCF values between LOS+LOS and LOS+NLOS conditions.

\begin{table}[ht]
\caption{The value of PCF at each measurement position\cite{ref:陈文俊人体}}
\centering
\renewcommand{\arraystretch}{1.5}
\small
\tabcolsep=0.2cm
\begin{tabular}{c|c|c|c|c|c}
\hline
\hline
\textbf{Position} & \textbf{Condition} & $O_{\text{back}}$ & \textbf{Position} & \textbf{Condition} & $O_{\text{back}}$ \\ \hline
1 & LOS+LOS & 0.89 & 8  & LOS+LOS  & 0.78 \\ \hline
2 & LOS+LOS & 0.73 & 9  & LOS+LOS  & 0.91 \\ \hline
3 & LOS+LOS & 0.67 & 10 & LOS+LOS  & 0.93 \\ \hline
4 & LOS+LOS & 0.75 & 11 & LOS+NLOS & 0.89 \\ \hline
5 & LOS+LOS & 0.84 & 12 & LOS+NLOS & 0.90 \\ \hline
6 & LOS+LOS & 0.81 & 13 & LOS+NLOS & 0.92 \\ \hline
7 & LOS+LOS & 0.86 & 14 & LOS+NLOS & 0.95 \\ \hline
\hline
\end{tabular}
\label{tab:O因子具体数值}
\end{table}

\begin{figure}[ht]
    \centering
    \includegraphics[width=\columnwidth]{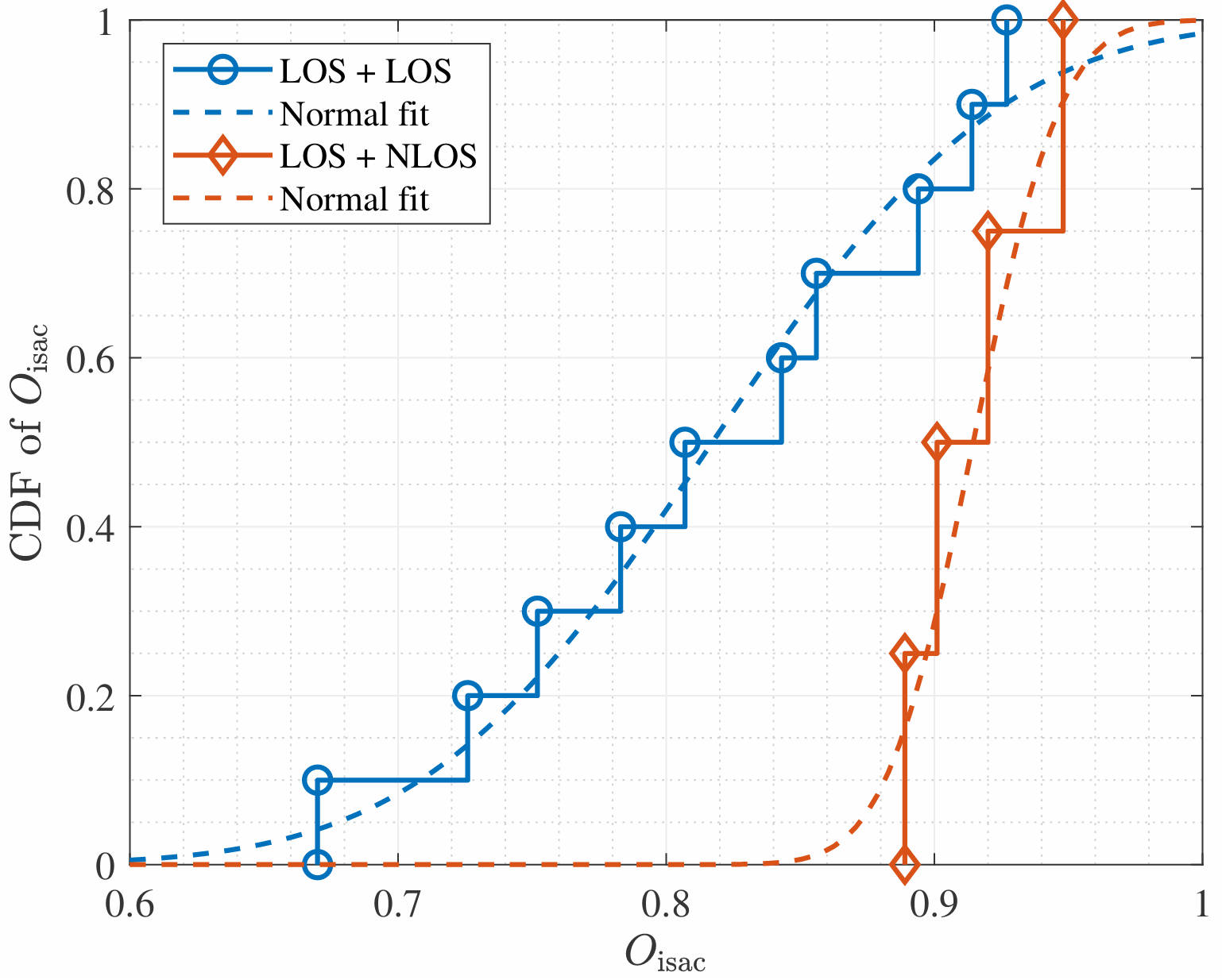}
    \caption{CDF of the PCF value $O_{\text{isac}}$. The blue and orange curves represent human target being in LOS+LOS and LOS+NLOS conditions, respectively\cite{ref:陈文俊人体}.}
    \label{fig:O因子CDF图}
\end{figure}

To further illustrate the distribution characteristics of PCF values, Fig.~\ref{fig:O因子CDF图} presents the cumulative distribution function (CDF) of PCF values under LOS+LOS and LOS+NLOS conditions, along with their normal distribution fits. The figure shows a distinct separation in the numerical distribution of PCF values between LOS+LOS and LOS+NLOS conditions, indicating that the position and LOS condition of the human target significantly impact channel path loss. Ultimately, the PCF was integrated into the ISAC path loss model, allowing the new model to more accurately reflect the combined effects of target and background channels, providing a more precise basis for modeling path loss in ISAC channels affected by human interference.

\subsection{Coupling properties between background channels}
\begin{figure}[!htbp]
	\centering
	\subfloat[]{\includegraphics[width=0.5\columnwidth]{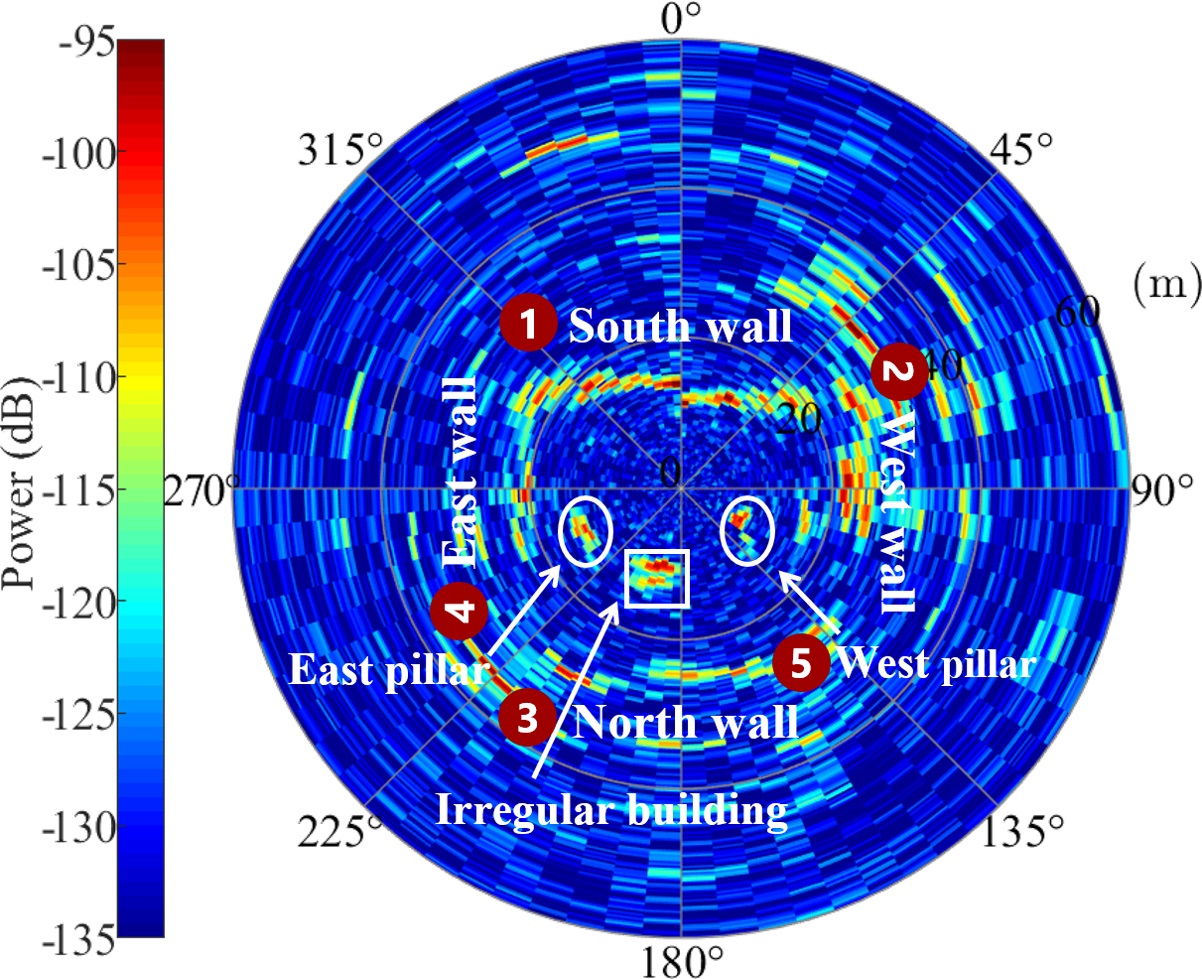}}
	\hfill
	\subfloat[]{\includegraphics[width=0.5\columnwidth]{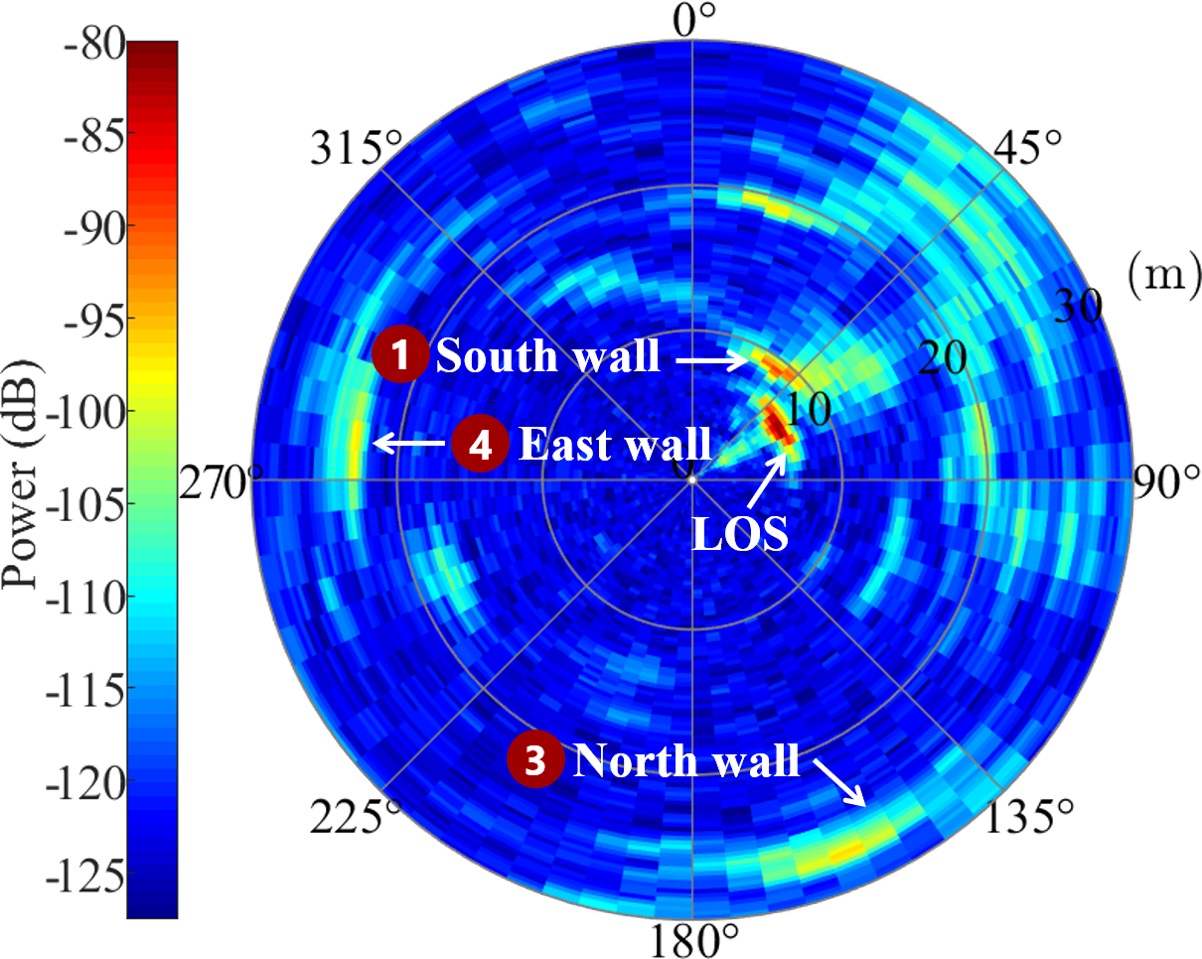}}
	\caption{Channel measurement for ISAC sharing feature.
		(a) mono-static PADP; (b) bi-static PADP\cite{ref:刘亚萌共享簇}.}
	\label{figs:室内大厅PADP图}
\end{figure}

In the study of the coupling relationship between the background channel and the target channel, we found that due to experiencing similar propagation environments, their large-scale path loss is influenced by the target's position. Similarly, for the same scenario, the sensing results of the mono-static and bi-static setups also exhibit a certain degree of correlation. To analyze the coupling relationship between the background channels, we further conducted measurements in Scen7: Indoor hall.
From the measurement results, it is observed that the mono-static sensing PADP shown in Fig.~\ref{figs:室内大厅PADP图}(a) can reconstruct the real environment quite clearly, while the PADP in Fig.~\ref{figs:室内大厅PADP图}(b) reflects the main propagation scatterers. Taking the mono-static sensing channel and one of the bi-static sensing channels as an example, by comparing Figs.~\ref{figs:室内大厅PADP图}(a) and~\ref{figs:室内大厅PADP图}(b), it can be seen that some environmental objects affect both the background and target links, acting not only as sensing targets but also as environmental scatterers for the communication channel. For instance, objects like the south wall (label 1), the north wall (label 3), and the east wall (label 4) in the figures demonstrate the existence of shared scatterers. Furthermore, we utilize a clustering algorithm\cite{ref:刘亚萌共享簇} to extract shared clusters generated by these shared scatterers and conduct a quantitative analysis of the number and distribution of these shared clusters. To measure the degree of sharing in the ISAC channels, we introduce a sharing degree (SD) parameter, defined as the proportion of power contributed by the shared clusters in the channel. The SD for the communication channel is expressed as follows.

\begin{equation}
	SD = \frac{P_{\text{shared}}^{\text{back}}}{P_{\text{total}}^{\text{back}}} 
	= \frac{\left|\sum_{n_0} \sum_{m_0} a_{\text{back},n_0,m_0} \sigma_{n_0,m_0}\right|^2}
	{\left|\sum_{n_0} \sum_{m_0} a_{\text{back},n_0,m_0} \sigma_{n_0,m_0} + \sum_{n_1} \sum_{m_1} a_{\text{back},n_1,m_1} \sigma_{n_1,m_1} \right|^2},
\end{equation}

In the equation, \(n_0\), \(m_0\), and \(n_1\), \(m_1\) represent the shared and non-shared clusters and paths in the background channel. \(a\) and \(\sigma\) represent the complex gain and scattering gain of the paths. \(P_{\text{shared}}^{\text{back}}\) and \(P_{\text{total}}^{\text{back}}\) represent the power of shared clusters and all clusters in the communication channel, respectively. Similarly, the background channel’s SD is defined as \(SD_{\text{back}} = {P_{\text{shared}}^{\text{back}}}/{P_{\text{total}}^{\text{back}}}\). In these equations, the ratio \(P_{\text{shared}} / P_{\text{total}}\) indicates the proportion of the shared part in the total received power of each channel. The SD value of the ISAC channel is related to the number of shared clusters, meaning that the higher the proportion of shared clusters, the higher the SD value.

\section{Conclusion}
This article presents a comprehensive study on the modeling and experimental validation of ISAC channels. 
Firstly, we propose an E-GBSM for ISAC channel modeling. The ISAC channel modeling can be divided into two parts: target and background channel modeling. For the target channel, we develop a concatenated modeling method through the convolution of Tx-ST and ST links. For the background channel, we consider its correlation with the target channel and develop a model applicable to both mono-static and bi-static sensing modes. 
Secondly, we conducted extensive channel measurements in various indoor and outdoor scenarios, covering diverse sensing targets such as metal plates, RIS, human bodies, UAVs, and vehicles. The results show that the Tx-ST-Rx links of the target channel can be constructed by convolving the Tx-ST and ST-link, and the significant power contribution of NLOS paths in these links necessitates their modeling. Furthermore, measurements reveal shared scatterers between the target and background channels, prompting us to incorporate the PCF to characterize the impact of the ST on the background channel. This factor can be statistically generated and utilized for the joint design of background and target channel models.

For future study, greater emphasis should be placed on the impact of ST properties such as RCS on the target channel. Additionally, it is necessary to explore how the correlation between the target and background channels can be leveraged to refine the ISAC channel model. Finally, extensive field measurements across various scenarios are required to validate the ISAC channel model and calibrate its parameters.

\section*{Acknowledgements}
This research is supported by Young Scientists Fund of the National Natural Science Foundation of China (62201087, 62101069), National Key R\&D Program of China (2023YFB2904803), Guangdong Major Project of Basic and Applied Basic Research (2023B0303000001), Key Program of National Natural Science Foundation of China (92167202), National Science Fund for Distinguished Young Scholars (61925102), National Natural Science Foundation of China (62341128) and Beijing University ot Posts and Telecommunications-China Mobile Research Institute Joint innovation Center.

\bibliographystyle{elsarticle-num}
\balance
\bibliography{egbib}

\begin{thebibliography}{10}
\expandafter\ifx\csname url\endcsname\relax
  \def\url#1{\texttt{#1}}\fi
\expandafter\ifx\csname urlprefix\endcsname\relax\def\urlprefix{URL }\fi
\expandafter\ifx\csname href\endcsname\relax
  \def\href#1#2{#2} \def\path#1{#1}\fi

\bibitem{ref:ITU愿景}
{International Telecommunication Union (ITU)}, {Future Technology Trends of Terrestrial International Mobile Telecommunications Systems Towards 2030 and Beyond}, available: \url{https://www.itu.int/dms_pub/itu-r/opb/rep/R-REP-M.2516-2022-PDF-E.pdf} (November 2022).

\bibitem{ref:ISAC综述}
Z.~Wei, F.~Liu, C.~Masouros, N.~Su, A.~P. Petropulu, Toward multi-functional 6g wireless networks: integrating sensing, communication, and security, IEEE Communications Magazine 60~(4) (2022) 65--71.

\bibitem{ref:需要的信道模型综述2}
Y.~Li, J.~Zhang, Z.~Ma, Y.~Zhang, Clustering analysis in the wireless propagation channel with a variational {Gaussian} mixture model, IEEE Transactions on Big Data 6~(2) (2018) 223--232.

\bibitem{ref:统计簇调研}
J.~Zhang, J.~Lin, P.~Tang, Y.~Zhang, H.~Xu, T.~Gao, H.~Miao, Z.~Chai, Z.~Zhou, Y.~Li, et~al., Channel measurement, modeling, and simulation for 6g: A survey and tutorial, arXiv preprint arXiv:2305.16616.

\bibitem{jiang2024high}
H.~Jiang, W.~Shi, X.~Chen, Q.~Zhu, Z.~Chen, High-efficient near-field channel characteristics analysis for large-scale mimo communication systems, IEEE Internet of Things Journal.

\bibitem{ref:TR38.901}
{3rd Generation Partnership Project (3GPP) RAN}, {Study on channel model for frequencies from 0.5 to 100 GHz}, Tech. Rep. {TR 38.901}, {3GPP}, available: \url{https://portal.3gpp.org/desktopmodules/Specifications/SpecificationDetails.aspx?specificationId=3173} (2019).

\bibitem{ref:TR22.137}
{3rd Generation Partnership Project (3GPP)}, {Integrated Sensing and Communication (ISAC)}, Tech. Rep. TS 22.137, 3GPP, available: \url{https://portal.3gpp.org/desktopmodules/Specifications/SpecificationDetails.aspx?specificationId=4198} (2022).

\bibitem{ref:116次会议}
{3rd Generation Partnership Project (3GPP)}, {Chair notes, 3GPP TSG RAN WG1 \#116, Meeting Report TSGR1\#116}, {3GPP, February 26th -- March 4th, 2024}, available: \url{https://www.3gpp.org/ftp/tsg_ran/WG1_RL1/TSGR1_116/Inbox/Chair_notes} (Mar. 2024).

\bibitem{ref:116b次会议}
{3rd Generation Partnership Project (3GPP)}, {Chair notes, 3GPP TSG RAN WG1 \#116b, Meeting Report TSGR1\#116b}, {3GPP, March 18th -- March 22nd, 2024}, available: \url{https://www.3gpp.org/ftp/tsg_ran/WG1_RL1/TSGR1_116b/Inbox/Chair_notes} (Mar. 2024).

\bibitem{ref:117次会议}
{3rd Generation Partnership Project (3GPP)}, {Chair notes, 3GPP TSG RAN WG1 \#117, Meeting Report TSGR1\#117}, {3GPP, May 6th -- May 10th, 2024}, available: \url{https://www.3gpp.org/ftp/tsg_ran/WG1_RL1/TSGR1_117/Inbox/Chair_notes} (May 2024).

\bibitem{ref:118次会议}
{3rd Generation Partnership Project (3GPP)}, {Chair notes, 3GPP TSG RAN WG1 \#118, Meeting Report TSGR1\#118}, {3GPP, June 17th -- June 21st, 2024}, available: \url{https://www.3gpp.org/ftp/tsg_ran/WG1_RL1/TSGR1_118/Inbox/Chair_notes} (Jun. 2024).

\bibitem{ref:118b次会议}
{3rd Generation Partnership Project (3GPP)}, {Chair notes, 3GPP TSG RAN WG1 \#118b, Meeting Report TSGR1\#118b}, {3GPP, October 14th -- October 18th, 2024}, available: \url{https://www.3gpp.org/ftp/tsg_ran/WG1_RL1/TSGR1_118b/Inbox/Chair_notes} (Oct. 2024).

\bibitem{ref:雷达相关调研}
Z.~Zhang, R.~He, B.~Ai, M.~Yang, X.~Zhang, Z.~Qi, Y.~Yuan, Channel measurements and modeling for dynamic vehicular isac scenarios at 28 ghz, arXiv preprint arXiv:2403.00605.

\bibitem{ref:裴元鹏news}
Y.~Zhang, J.~Zhang, Y.~Pei, Y.~Liu, T.~Jiang, Latest progress for 3gpp isac channel modeling standardization, Science China(Information Sciences) 67~(11) (2024) 357--358.

\bibitem{ref:北交综述和簇替换}
Z.~Zhang, R.~He, B.~Ai, M.~Yang, C.~Li, H.~Mi, Z.~Zhang, A general channel model for integrated sensing and communication scenarios, IEEE Communications Magazine 61~(5) (2022) 68--74.

\bibitem{ref:Oppo提案}
OPPO, R1-2400617: Study on isac channel modelling, 3GPP TSG RAN WG1 \#116, Agenda Item 9.7.2, Document for Discussion and Decision, meeting held February 26th -- March 4th, 2024. Available: \url{https://www.3gpp.org/ftp/tsg_ran/WG1_RL1/TSGR1_116/Docs/R1-2400617%20Study%20on%20ISAC%20channel%20modelling.docx} (Feb. 2024).

\bibitem{ref:理论分析级联建模}
D.~Arnitz, U.~Muehlmann, K.~Witrisal, Wideband characterization of backscatter channels: Derivations and theoretical background, IEEE Transactions on Antennas and Propagation 60~(1) (2011) 257--266.

\bibitem{ref:背景调研3}
J.~Lou, R.~Liu, C.~Jiang, X.~Han, Z.~Han, Q.~Yang, Z.~Wang, A unified channel model for both communication and sensing in integrated sensing and communication systems, in: 2023 IEEE 98th Vehicular Technology Conference (VTC2023-Fall), IEEE, 2023, pp. 1--6.

\bibitem{ref:李3}
X.~Li, Q.~Wang, M.~Zeng, Y.~Liu, S.~Dang, T.~A. Tsiftsis, O.~A. Dobre, Physical-layer authentication for ambient backscatter-aided noma symbiotic systems, IEEE Transactions on Communications 71~(4) (2023) 2288--2303.

\bibitem{ref:单站与雷达关系的调研}
S.~Lu, F.~Liu, L.~Hanzo, The degrees-of-freedom in monostatic isac channels: Nlos exploitation vs. reduction, IEEE Transactions on Vehicular Technology 72~(2) (2023) 2643--2648.
\newblock \href {http://dx.doi.org/10.1109/TVT.2022.3210307} {\path{doi:10.1109/TVT.2022.3210307}}.

\bibitem{ref:共享簇调研3}
A.~Ali, N.~Gonz{\'a}lez-Prelcic, A.~Ghosh, Passive radar at the roadside unit to configure millimeter wave vehicle-to-infrastructure links, IEEE Transactions on Vehicular Technology 69~(12) (2020) 14903--14917.

\bibitem{ref:共享簇调研5}
T.~T. Nguyen, K.~Elbassioni, N.~C. Luong, D.~Niyato, D.~I. Kim, Access management in joint sensing and communication systems: Efficiency versus fairness, IEEE Transactions on Vehicular Technology 71~(5) (2022) 5128--5142.

\bibitem{ref:共享簇调研6}
M.~L. Rahman, J.~A. Zhang, X.~Huang, Y.~J. Guo, R.~W. Heath, Framework for a perceptive mobile network using joint communication and radar sensing, IEEE Transactions on Aerospace and Electronic Systems 56~(3) (2019) 1926--1941.

\bibitem{ref:共享簇调研2}
X.~Chen, Z.~Feng, Z.~Wei, P.~Zhang, X.~Yuan, Code-division ofdm joint communication and sensing system for 6g machine-type communication, IEEE Internet of Things Journal 8~(15) (2021) 12093--12105.

\bibitem{ref:共享簇调研4}
A.~L{\'o}pez-Reche, D.~Prado-Alvarez, A.~Ramos, S.~Inca, J.~F. Monserrat, Y.~Zhang, Z.~Yu, Y.~Chen, Considering correlation between sensed and communication channels in gbsm for 6g isac applications, in: 2022 IEEE Globecom Workshops (GC Wkshps), IEEE, 2022, pp. 1317--1322.

\bibitem{ref:3DMIMO的JSAC}
J.~Zhang, Y.~Zhang, Y.~Yu, R.~Xu, Q.~Zheng, P.~Zhang, 3-d mimo: How much does it meet our expectations observed from channel measurements?, IEEE Journal on Selected Areas in Communications 35~(8) (2017) 1887--1903.

\bibitem{ref:需要的信道模型综述1}
J.~Zhang, C.~Pan, F.~Pei, G.~Liu, X.~Cheng, Three-dimensional fading channel models: A survey of elevation angle research, IEEE communications magazine 52~(6) (2014) 218--226.

\bibitem{ref:陈文俊人体}
W.~Chen, Y.~Zhang, Y.~Liu, J.~Zhang, H.~Gong, T.~Jiang, L.~Xia, An empirical study of isac channel characteristics with human target impact at 105 ghz, Electronics Letters 60~(17) (2024) e70017.

\bibitem{ref:李1}
X.~Li, M.~Zhang, H.~Chen, C.~Han, L.~Li, D.-T. Do, S.~Mumtaz, A.~Nallanathan, Uav-enabled multi-pair massive mimo-noma relay systems with low-resolution adcs/dacs, IEEE Transactions on Vehicular Technology.

\bibitem{ref:李2}
M.~Deng, Z.~Yao, X.~Li, H.~Wang, A.~Nallanathan, Z.~Zhang, Dynamic multi-objective awpso in dt-assisted uav cooperative task assignment, IEEE Journal on Selected Areas in Communications 41~(11) (2023) 3444--3460.
\newblock \href {http://dx.doi.org/10.1109/JSAC.2023.3310056} {\path{doi:10.1109/JSAC.2023.3310056}}.

\bibitem{ref:张骥威RIS级联}
J.~Zhang, Y.~Zhang, T.~Jiang, H.~Gong, H.~Xing, L.~Tian, Cascaded channel modeling and experimental validation for ris assisted communication system, arXiv preprint arXiv:2412.07356.

\bibitem{ref:李毅波导}
Y.~Li, J.~Zhang, P.~Tang, L.~Tian, X.~Zhao, H.~Xu, H.~Gong, Path loss modeling for the ris-assisted channel in a corridor scenario in mmwave bands, in: 2022 IEEE Globecom Workshops (GC Wkshps), IEEE, 2022, pp. 1478--1483.

\bibitem{ref:周正福RIS}
J.~Zhang, Z.~Zhou, Y.~Zhang, L.~Tian, Z.~Yuan, T.~Jiang, A deterministic channel modeling method for ris-assisted communication in sub-thz frequencies, in: Accept by 2023 17th European Conference on Antennas and Propagation (EuCAP), 2023.

\bibitem{ref:王嘉琳测量}
J.~Wang, J.~Zhang, Y.~Zhang, T.~Jiang, L.~Yu, G.~Liu, Empirical analysis of sensing channel characteristics and environment effects at 28 ghz, in: 2022 IEEE Globecom Workshops (GC Wkshps), IEEE, 2022, pp. 1323--1328.

\bibitem{ref:刘亚萌共享簇}
Y.~Liu, J.~Zhang, Y.~Zhang, Z.~Yuan, G.~Liu, A shared cluster-based stochastic channel model for integrated sensing and communication systems, IEEE Transactions on Vehicular Technology.

\end{thebibliography}

\end{document}